\documentclass[12pt,a4paper]{article}
\usepackage{booktabs,latexsym}
\usepackage{graphicx,cite,amsmath,times,subfigure,epsfig,lscape,longtable}
\usepackage{txfonts}
\usepackage{times,mathptm}
\usepackage{amssymb}

\title{\Large An optimized dispersion-relation-preserving combined compact difference scheme to solve advection equations}
\author
{\date{}%\itshape
C. H. Yu$^{1}$,
D. Wang$^{1}$,
Z. He$^{1,}$\thanks{\scriptsize
Corresponding author. Tel:+86-571-88208912; Fax:+86-571-88208890.
E-mail address: hezhiguo@zju.edu.cn (Z. He)}
,~T. P\"ahtz$^{1}$
\\ \vspace{-1mm} \small
\itshape $^{1}$Ocean College, Zhejiang University, 866 Yuhangtang Road,
\\ \small \itshape Hangzhou, Zhejiang, People's Republic of China}

\begin{document}
\maketitle

\abstract{
%The dispersion relation of a hyperbolic partial differential equations (PDE) relates the phase velocity to the wavelength of waves solving of the PDE.
%A dispersion relation of a advection equation relates frequency to wavenumber of a wave.
%For a numerical scheme solving a advection equation,
%it is crucial to preserve this dispersion relation.

In this study,
we first present an improved version of the classical sixth-order combined compact difference (CCD6) scheme
to enhance the convective stability of advection equations through an increased dispersion accuracy.
This improved fifth-order dispersion-relation-preserving combined compact difference scheme (DRPCCD5)
has been rigorously analyzed through the dispersion, phase speed anisotropy and stability analyses.
We then couple the DRPCCD5 scheme with the previous fifth-order compact-reconstruction
weighted essentially non-oscillatory (CRWENO5) scheme using a novel hybrid strategy based on the monotonicity-maintenance criteria.
To verify the resulting "optimized" hybrid scheme (ODRPCCD5), several benchmark problems with available exact solution are investigated.
The comparison to the previous fifth-order WENO (WENO5) scheme shows that the ODRPCCD5 avoids numerical oscillation around discontinuities, handles large gradients well,
and is much faster at the same accuracy because a coarser mesh can be used.

%A property of a numerical scheme, which  is crucial for the simulation of flow fields
%which involve different length scales.
%In this paper,
%a fifth-order dispersion-relation-preserving combined compact difference (DRPCCD5) scheme,
%which accommodates a better dispersion relation in the hyperbolic conservation laws,
%is proposed to enhance the convective stability by virtue of the increased dispersive accuracy.
%This DRPCCD5 scheme has been rigorously developed
%within the four-stencil point framework through the dispersion and dissipation analysis.
%Furthermore,
%DRPCCD5 scheme is coupled with an fifth-order compact-reconstruction
%weighted essentially non-oscillatory (CRWENO5) scheme to make it possible for the discontinuity-capturing.
%In other words,
%an DRPCCD5 scheme has the desired dispersion property and thus,
%coupled with a non-oscillatory limiter,
%are ideal simulation for the numerical solution of hyperbolic conservation laws.
%To verify the proposed hybrid  DRPCCD5/CRWENO5 scheme,
%several problems that are amenable to the exact solutions will be investigated.
%Comparing the proposed scheme with fifth-order weighted essentially non-oscillatory (WENO5)
%scheme shows that our method reduces smearing near shocks and corners,
%and it is more accurate near high gradient region.

\begin{description}
\item[{\footnotesize\bf Keywords:}]
{\footnotesize }
combined compact difference scheme;
dispersion-relation-preserving;
non-oscillatory;
monotonicity-maintenance criteria;
large gradients.
\end{description}

\clearpage

\section{Introduction}
\ \ \ \ \
%The dispersion relation of a hyperbolic partial differential equations (PDE),
%which can be derived from the spatial Fourier transform of the PDE, relates the phase velocity to the wavelength of waves solving of the PDE.
Numerical simulations of advection equations are commonly found in many applications of practical importance, such as shock waves, shallow water flow,
magnetohydrodynamics, and two-phase flow models. When numerically solving such convection-dominated partial differential equations (PDEs),
it is desirable to minimize the indispensable dispersion error,
which is defined as the discrepancy between the numerical and actual wavenumbers,
because this enhances convective stability and allows for accurate capturing of small length scales in the wave phase \cite{bib:Chiu228(2009)3640-3655}.

For this purpose,
dispersion-relation-preserving (DRP) approaches have been developed to enhance convective stability by rigorously preserving the dispersion relation
\cite{bib:Tam107(1993)262-281,bib:Wang174(2001)381-404,bib:Popescu2210(2005)705-729,bib:Chiu228(2009)3640-3655,bib:Chiu228(2009)8034-8052,bib:Bhumkar278(2014)378-399}.
Furthermore,
compact difference schemes offer spectral accuracy with fewer grid points to improve convective stability
\cite{bib:Lele103(1992)16-24,bib:Sengupta192(2003)667-694,bib:Sengupta26(2006)151-193,bib:Wilson39(3)(2001)}.
These compact difference schemes have been extended to combined compact difference schemes (CCD) \cite{bib:Chu140(1998)370-399},
in which first and second derivative terms
are simultaneously evaluated in an implicit manner, making the scheme more compact and accurate.
%Although compact schemes and CCD schemes in comparison to explicit finite difference schemes
%not only offer higher order approximations to differential operators
%but also provide higher resolution with the same girds points.
%However, the main drawback of these compact schemes are that the spectral properties are not good enough.
CCD schemes suffer from stability issues of boundary conditions when solving the PDE.
In fact, these schemes need special treatment at the boundary nodes, in particular when simulating thin boundary layer problems.
Hence, the boundary closures have been improved \cite{bib:Sengupta228(2009)3048-3071} to obtain better numerical properties,
and the corresponding dissipation and de-aliasing properties have been discussed \cite{bib:Sengupta228(2009)6150-6168}.
%Alternatively, dispersion relation-preserving (DRP) approaches enhance convective stability by rigorously preserving the dispersion relation
%\cite{bib:Tam107(1993)262-281,bib:Wang174(2001)381-404,bib:Popescu2210(2005)705-729,bib:Chiu228(2009)3640-3655,bib:Chiu228(2009)8034-8052,bib:Bhumkar278(2014)378-399}.

High spectral resolution schemes,
such as the compact difference and CCD schemes,
inevitably produce numerical oscillations near discontinuities
and lead to failure of the flow simulation.
In order to avoid numerical oscillations,
high resolution schemes often use flux/slope limiters to bound the solution gradient
around shocks or discontinuities \cite{bib:Harten49(1983),bib:Leonard88(1991)}.
Some representative schemes belonging to this class of methods
include the essentially non-oscillatory (ENO) scheme \cite{bib:Shu77(1988)439-471,bib:Shu83(1989)32-78} and their weighted variants,
known as the weighted ENO (WENO) \cite{bib:Lin1(1994)200-212,bib:Jiang126(1996)202-228}.
It's well known that the ENO and WENO schemes may be too dissipative for compressible turbulence simulations and aero-acoustics problems.
Hence, the compact-reconstruction weighted essentially non-oscillatory (CRWENO) scheme
\cite{bib:Ghosh34(3)(2012)} has been presented,
in which compact sub-stencils are identified at each interface and combined using the WENO weights.
WENO schemes have been intensively used for problems
containing both shocks and complicated smooth solution structures \cite{bib:Martin220(2006)270-289,bib:Shyuea268(2014)326-354}.

Algorithms with high accuracy are required to capture small wavelengths and non-oscillatory behaviors across discontinuities like shock waves.
For this purpose, special finite difference schemes have been introduced \cite{bib:Suna230(12)(2011)4616-4635}. Also, the hybrid finite difference scheme based
on the minimized dispersion and controllable dissipation (MDCD) technique has been developed to solve advection equations.
This MDCD technique has been coupled with an optimized WENO scheme to make discontinuity capturing possible \cite{bib:Suna270(2014)238-254}.
%While the hybrid schemes constructed show improved resolution
%while being non-oscillatory, they require indicators to detect discontinuities.
%They also suffer from the drawback that they switch to a non-compact scheme at and near discontinuities,
%thus resulting in a loss of resolution.
Many researchers have also proposed various alternative ways to improve the numerical schemes
\cite{bib:Adams127(1996)27-51,bib:Pirozzoli178(2002)81-117,bib:Ren192(2003)365-386,bib:Zhou227(2007)}.
However, accuracy still remains a challenge because,
to our knowledge,
most if not all existing numerical schemes suffer from the drawback that they switch to a non-compact scheme at and near discontinuities,
resulting in a loss of resolution.

In this study,
a fifth-order dispersion-relation-preserving combined compact difference (DRPCCD5) scheme
which has better DRP properties than previously reported compact difference schemes over a considerable range of wavenumbers is proposed.
This scheme ensures that resolved energy components propagate closer to the correct physical speed,
and that complex phenomena, involving interactions among different wavelength scales, can be captured.
%The present DRPCCD5 is rigorously analyzed through fundamental analysis
%in comparison with other two existed explicit finite difference scheme and compact difference scheme.
Furthermore,
the DRPCCD5 scheme is coupled with the CRWENO5 scheme using a novel hybrid strategy based on the monotonicity-maintenance criteria. The numerical properties of the
resulting "optimized" hybrid scheme (ODRPCCD5) are then rigorously analyzed using several benchmark problems.

%the previous explicit finite difference scheme,
%compact difference schemes
%and present dispersion-relation-preserving combined compact difference (DRPCCD5) are rigorously analyzed through fundamental analysis.
%We find that the compact difference scheme in comparison to the explicit finite difference scheme
%not only offer higher order approximations to differential operators but also reduces dispersion errors.
%However, the main drawback of compact difference scheme is that the DRP properties are not optimal.}
%A DRPCCD5 scheme which has better DRP properties than previously reported compact difference scheme over a considerable range of wavenumbers was proposed.
%This ensures that resolved energy components propagate closer to the correct physical speed,
%and that complex phenomena which involve interactions among different wavelength scales can be captured.
%Furthermore,
%this DRPCCD5 scheme is coupled with an fifth-order compact-reconstruction weighted essentially non-oscillation (CRWENO5) scheme
%to make discontinuity capturing possible by {\color{red}hybridized strategy depended on Courant number,
%which is different from a spectral-optimized combined compact scheme
%hybridized with WENO strategy \cite{bib:Zhou227(2007)}.}

This paper is organized as follows.
Section 2 describes discretization of a standard advection equation
and the time marching method, which is used in the present study.
The schemes construction is carried out in Section 3.
Section 4 includes the fundamental analysis of dispersion, dissipation, phase speed anisotropy,
numerical group velocity, and numerical phase velocity for the proposed DRPCCD5 scheme.
Several benchmark tests are performed in Section 5 to validate the ODRPCCD5 scheme.
Section 6 draws concluding remarks based on the results presented in Section 5.

\section{Time marching method}
\ \ \
The one-dimensional linear wave equation can be expressed as
\begin{align}
\label{eq:1dwave}
\frac{\partial u}{\partial t}+ \frac{\partial f}{\partial x}=0.
\end{align}
where $t$ is time, $x$ the spatial coordinate, $u$ the field variable, and $f=c u$ with $c$ the constant propagation speed of the wave.
%Integrating Eq. (\ref{eq:1dwave}) over the cell $[x_{i-\frac{1}{2}},x_{i+\frac{1}{2}}]$
%leads to the semi-discrete conservation equation
A conservative finite difference discretization of Eq.~(\ref{eq:1dwave}) results in an ordinary
differential equation, which can be expressed as %\cite{bib:Jiang126(1996)202-228,bib:Suna270(2014)238-254}
%A conservative finite difference discretization of (\ref{eq:1dwave}) on a uniform grid is
\begin{align}
\label{eq:1dwave2}
\frac{d u_{i}}{d t} =  F_{i}(u)= -\frac{1}{h} (\hat{f}_{i+\frac{1}{2}} - \hat{f}_{i-\frac{1}{2}} ).
\end{align}
%where $h$ is the grid spacing and $\hat{f}_{i+\frac{1}{2}}$ is the interfacial value of $f$ between $x_i$ and $x_{i+1}$.
where $h$ is the grid spacing and $\hat{f}_{i+\frac{1}{2}}$ is the numerical approximation of flux between points  $x_i$ and $x_{i+1}$.
In the present study,
we apply the fourth-order Runge-Kutta (RK4) scheme and the
sixth-order symplectic Runge-Kutta (SRK6) scheme \cite{bib:Oevel(1997)} for time evolution.
The explicit RK4 scheme reads
\begin{align}
\label{eq:nonTVD R-K}
u^{(1)}&=u^{(0)}+\frac{\Delta t}{2}~F(u^{(0)}),\notag \\
u^{(2)}&=u^{(1)}+\frac{\Delta t}{2}~(-F(u^{(0)})+F(u^{(1)})),\notag \\
u^{(3)}&=u^{(2)}+\frac{\Delta t}{2}~(-F(u^{(1)})+2F(u^{(2)})),\notag \\
u^{(4)}&=u^{(3)}+\frac{\Delta t}{6}~(F(u^{(0)})+2F(u^{(1)})-4F(u^{(2)})+F(u^{(3)})).
\end{align}
For the SRK6 scheme,
given the solution $u^{n}$ at $t=n\Delta t$,
the solution $u^{n+1}$ is obtained from the following iteration.
We start with computing $u^{(j)}$ and $F^{(j)}=F(u^{(j)})$,
where $j$=1 to 3, by numerically solving the following equations iteratively:
\begin{align}
\label{eq:sym-1}
u^{(1)}=u^{n}+\Delta t~[\frac{5}{36}F^{(1)}+(\frac{2}{9}+\frac{2\widetilde{c}}{3})F^{(2)}
+\frac{5}{36}+\frac{\widetilde{c}}{3})F^{(3)}],
\end{align}
\begin{align}
\label{eq:sym-2}
u^{(2)}=u^{n}+\Delta t~[(\frac{5}{36}-\frac{5\widetilde{c}}{12})F^{(1)}+\frac{2}{9}F^{(2)}
+(\frac{5}{36}+\frac{5\widetilde{c}}{12})F^{(3)}],
\end{align}
\begin{align}
\label{eq:sym-3}
u^{(3)}=u^{n}+\Delta t~[(\frac{5}{36}-\frac{\widetilde{c}}{3})F^{(1)}+(\frac{2}{9}-\frac{2\widetilde{c}}{3})F^{(2)}
+\frac{5}{36}F^{(3)}],
\end{align}
where $\widetilde{c}=\frac{1}{2}\sqrt{\frac{3}{5}}$. These updated values correspond to the times $t=n+(\frac{1}{2}+\widetilde{c})\Delta t$,
$t=n+\frac{1}{2}\Delta t$, and $t=n+(\frac{1}{2}-\widetilde{c})\Delta t$, respectively.
Upon reaching the user's specified tolerance ($10^{-8}$),
the solution at $t=(n+1)\Delta t$ is obtained as
\begin{align}
\label{eq:sym-4}
u^{n+1}=u^{n}+\frac{\Delta t}{9}~[\frac{5}{2}F^{(1)}+4F^{(2)}+\frac{5}{2}F^{(3)}].
\end{align}
The RK4 scheme is mainly used to run the numerical tests in this study
because the implicit SRK6 scheme provides nearly the same results, but is very time-consuming
(see results of linear advection problem \#1 in Table 1 and Fig. 7).

\section{Numerical Schemes for spatial discretization}

%\subsection{Reconstruction of non-compact interpolations}
\subsection{Fifth-order non-compact difference scheme}
\ \ \ \
The numerical flux can be reconstructed using a left or right biased interpolation \cite{bib:Ghosh34(3)(2012)}.
The appropriate interpolation is chosen based on the sign of the wave speed,
which in the case of a scalar PDE is given by
%\begin{align}
%\label{eq:re-con}
%h_{i+1/2}&=h^+_{i+1/2}~~\mathbf{if} \;f^{'}(u)|_{x=x_{i+1/2}}>0, \notag\\
%&=h^-_{i+1/2}~~\mathbf{if} \;f^{'}(u)|_{x=x_{i+1/2}}<0.
%\end{align}
%\begin{align}
%\label{eq:re-con}
%\hat{f}_{i+1/2}&=\hat{f}^+_{i+1/2}~~\mathbf{if} \;f^{'}(u)|_{x=x_{i+1/2}}\geq0, \notag\\
%&=\hat{f}^-_{i+1/2}~~\mathbf{if} \;f^{'}(u)|_{x=x_{i+1/2}}<0.
%\end{align}
\begin{align}
\label{eq:re-con}
\hat{u}_{i+1/2}&=\hat{u}^+_{i+1/2}~~\mathrm{if}~\;c_{i+1/2}\geq0, \notag\\
&=\hat{u}^-_{i+1/2}~~\mathrm{if}~\;c_{i+1/2}<0.
\end{align}
where the superscripts $+$ and $-$ denote left and right biased interpolations respectively.
Note that the approximation of the left biased numerical $\hat{u}_{i+1/2}$ is described in this section.
%The numerical approximation of $h^{-}_{i+1/2}$ can be similarly derived.
%The reconstruction process seeks to approximate the numerical
The first derivative term $\frac{\partial u}{\partial x}$ can be approximated to the desired order ($r$), reading
\begin{align}
\label{eq:re-con1}
\frac{\partial u}{\partial x}|_{x=x_{j}}= \frac{\hat{u}_{i+1/2}-\hat{u}_{i-1/2}}{h}+O(h^{r}),
\end{align}
where $\hat{u}_{i+1/2}$ for odd $r$ is computed using the linear reconstruction on a stencil
\begin{align}
\label{eq:re-con2}
\hat{u}_{i+1/2}=\sum_{k=-(r-1)/2}^{(r-1)/2}~b_{k}u_{i+k}.
\end{align}
Here $b_{k}$ is the coefficient and $u_{i+k}= u_{x_{i}+k h}$.
For $r=5$,
it reads
%\begin{align}
%\label{eq:re-con3}
%\frac{3}{10}\hat{f}_{i-1/2}+\frac{6}{10}\hat{f}_{i+1/2}+\frac{1}{10}\hat{f}_{i+3/2}=
%\frac{1}{30}f_{i-1}+\frac{19}{30}f_{i}+\frac{10}{30}f_{j+1}
%\end{align}
\begin{align}
\label{eq:re-con3}
\hat{u}_{i+1/2} = \frac{1}{30}u_{i-2}-\frac{13}{60}u_{i-1}+\frac{47}{60}u_{i}+\frac{27}{60}u_{i+1}-\frac{1}{20}u_{i+2}.
\end{align}
This scheme has fifth-order spatial accuracy
according to the derived modified equation given below
\begin{align}%\nonumber
\label{eq:order2}
    \frac{\partial u}{\partial x} = \frac{\partial u}{\partial x}|_{exact}
  & +\frac{1}{60}~{h}^{5}\: \frac{\partial^{6} u}{\partial x^{6}}+ O(h^6)~.
\end{align}
%\begin{align}
%\label{eq:re-con4}
%\hat{f}_{i+1/2}=h_{i+1/2}-\frac{1}{600}\frac{\partial^{5} f}{\partial x^{5}}|_{x=x_i}{\Delta x^5}+O(\Delta x^{6})
%\end{align}
The drawback of this scheme is that the magnitude of the leading error of the resulting scheme is too large.
In addition,
this scheme must adopt a very fine mesh to correctly capture the important advection flow structures.
A space-time accurate numerical simulation of advection problems requires
higher spatial resolution and dispersion-relation-preserving (DRP) properties.
Such schemes act as an important numerical tool to solve complex physical problems displaying a large bandwidth of spatio-temporal scales.
For this reason,
we develop a new fifth-order dispersion-relation-preserving combined compact difference (DRPCCD5) scheme in the Section 3.2.

%\subsection{Reconstruction of compact interpolations for DRPCCD5 scheme}
%\ \ \ \
%The solution of a hyperbolic PDE  is composed of waves.
%A scalar solution comprises of a right or left running wave.
%The reconstruction of the numerical flux at the interfaces models this wave phenomenon through upwinding.
%At each interface,
%the flux can be approximated using a left or right biased interpolation.
%The appropriate interpolation is chosen based on the sign of the wave speed,
%which in the case of a scalar PDE is given by
%\begin{align}
%\label{eq:hea}
%h_{i+\frac{1}{2}}=\left\{
%\begin{array}{lcl}   %{cc}
%h^{L}_{i+\frac{1}{2}}    &~~ ; ~if~~f^{'}(u)|_{x=x_{i+\frac{1}{2}}}>0\\
%h^{R}_{i+\frac{1}{2}}     &   ;  if~~f^{'}(u)|_{x=x_{i+\frac{1}{2}}}>0~&.\\
%\end{array}
%\right.
%\end{align}
%where the superscripts $L$ and $R$ denote left and right biased interpolations respectively.
%The approximation of the left biased numerical flux $h^{L}_{i+\frac{1}{2}}$ is described in this section.
%The approximation of the right biased $h^{R}_{i+\frac{1}{2}}$ follows a similar process.
%The superscripts are omitted in the subsequent description.

\subsection{Fifth-order dispersion-relation-preserving combined compact difference (DRPCCD5) scheme}
\ \ \ \
%In the following the combined compact difference scheme for approximating
%the spatial derivative term $\frac{\partial f}{\partial x}$ is presented.
%Besides the derivative term $\frac{\partial f}{\partial x}$,
%the second derivative term $\frac{\partial^2 f}{\partial x^2}$ is considered as the unknown variable at each grid point
%so as to be able to get a spectral-like resolution.
In this section,
we present an improved upwind combined compact difference scheme.
The first and the second derivative terms ( $\frac{\partial u}{\partial x}$ and $\frac{\partial^2 u}{\partial x^2}$)
in a four-point grid stencil are approximated as
%In a four-point grid stencil,
%the numerical scheme for $\frac{\partial f}{\partial x}$ and
%$\frac{\partial^2 f}{\partial x^2}$ are given below
\begin{align}
\label{eq:CCD-EQ1-1}\nonumber
   a_1\frac{\partial u}{\partial x}|_{i-1}
     +\frac{\partial u}{\partial x}|_{i}
    +a_3\frac{\partial u}{\partial x}|_{i+1}\\
    =\frac{1}{h}( c_{1} u_{i-2} + c_{2}&u_{i-1} +c_{3}u_{i} )
      -h\left( b_1\frac{\partial^2 u}{\partial x^2}|_{i-1}
          +b_2\frac{\partial^2 u}{\partial x^2}|_{i}
          +b_3\frac{\partial^2 u}{\partial x^2}|_{i+1} \right), \\
   -\frac{1}{8}\frac{\partial^2 u}{\partial x^2}|_{i-1}
      +\frac{\partial^2 u}{\partial x^2}|_{i}
   -\frac{1}{8}\frac{\partial^2 u}{\partial x^2}|_{i+1}
    &=\frac{3}{h^2}( u_{i-1} -2u_{i} + u_{i+1} )
      -\frac{9}{8h}\left(-\frac{\partial u}{\partial x}|_{i-1}
     +                 \frac{\partial u}{\partial x}|_{i+1}\right).
   \label{eq:CCD-EQ2-1}
\end{align}

The coefficients shown in Eq.~(\ref{eq:CCD-EQ2-1}) are derived
through Taylor series expansion.
Elimination of the leading truncation error terms in the modified equation analysis enables us to get
the formal accuracy order of six \cite{bib:Chu140(1998)370-399}.

Derivation of the coefficients in Eq.~(\ref{eq:CCD-EQ1-1}) is started
from performing Taylor series expansion on the terms
$u_{i-2}$,
$u_{i-1}$,
$\frac{\partial u}{\partial x}|_{i-1}$,
$\frac{\partial u}{\partial x}|_{i}$,
$\frac{\partial u}{\partial x}|_{i+1}$,
$\frac{\partial^{2} u}{\partial x^{2}}|_{i-1}$,
$\frac{\partial^{2} u}{\partial x^{2}}|_{i}$
and
$\frac{\partial^{2} u}{\partial x^{2}}|_{i+1}$
with respect to $u_i$ to get the modified equation.
The six leading truncation error terms derived in the modified equation analysis
are then eliminated to get a set of six algebraic equations
\begin{align}
\label{eq:coff_eq1} c_1 + c_2 + c_3 = 0,  \\
\label{eq:coff_eq2} -2c_1-c_2-a_1-a_3=1, \\
\label{eq:coff_eq3} 4c_1+c_2+2a_1-2a_3-2b_1-2b_2-2b_3 = 0, \\
\label{eq:coff_eq4} 8c_1+c_2+3a_1+3a_3-6b_1+6b_3 = 0, \\
\label{eq:coff_eq5} 16c_1+c_2+4a_1-4a_3-12b_1-12b_3 = 0, \\
\label{eq:coff_eq6} 32c_1+c_2+5a_1+5a_3-20b_1+20b_3 = 0.
%\label{eq:coff_eq7} 64c_1+c_2+6a_1-6a_3-30b_1-30b_3 = 0.
\end{align}
Derivation of two further algebraic equations are needed to determine all eight coefficients in
Eq.~(\ref{eq:CCD-EQ1-1}).
%We are still short of two algebraic equations to uniquely determine all the introduced coefficients shown in Eq.~(\ref{eq:CCD-EQ1-1}).
One way of deriving the two equations so as to get a better approximation
of $\frac{\partial u}{\partial x}$ is to reduce numerical error of the accumulative type.
We can then expect to retain the theoretical dispersive property  of $\frac{\partial u}{\partial x}$\cite{bib:Tam107(1993)262-281}.

Our strategy of achieving the goal of reducing numerical dispersion error is to match the exact
and numerical wavenumbers.
Use of this underlying approach  amounts to equating the effective
wavenumbers  $\alpha^{'}$ and $\alpha^{''}$ to those shown on
the right-hand sides of Eqs.~(\ref{eq:transform3}) and (\ref{eq:transform4}) \cite{bib:Tam107(1993)262-281}.
Following this line of derivation,
we are led to get the two equations for $\alpha 'h$ and $\alpha ''h$ as follows
\begin{align}
\mathbf{i}\alpha^{'} h~( a_1 e^{-\mathbf{i}\alpha h} + 1 + a_3 e^{\mathbf{i}\alpha h })
=(c_1e^{-2\mathbf{i}\alpha h}+c_2e^{-\mathbf{i}\alpha h}+c_3)
- (\mathbf{i}\alpha^{''} h)^2(b_1e^{-\mathbf{i}\alpha h} + b_2 + b_3e^{\mathbf{i}\alpha h}),
\label{eq:transform3}
\end{align}
\begin{align}
(\mathbf{i}\alpha^{''} h)^2(-\frac{1}{8}e^{-\mathbf{i}\alpha h} +1-\frac{1}{8} e^{\mathbf{i}\alpha h})
=(3e^{-\mathbf{i}\alpha h}-6+3 e^{\mathbf{i}\alpha h})
-\mathbf{i}\alpha^{'} h~(-\frac{9}{8}e^{-\mathbf{i}\alpha h} +\frac{9}{8}e^{\mathbf{i}\alpha h}).
\label{eq:transform4}
\end{align}
Equations (\ref{eq:transform3}) and (\ref{eq:transform4}) are solved to get the expression for $\alpha^{'} h$ which has been used
subsequently to minimize the dispersion error.
The real and imaginary parts of $\alpha^{'} h$
provide information regarding the dispersion error (phase error) and dissipation error (amplitude error),
respectively.
%The expression of $\alpha' h$ can be directly derived from Eqs.~(\ref{eq:transform3}) and (\ref{eq:transform4}).
%It is worthy to note that the real and imaginary parts
%of the numerical modified (or scaled) wavenumber $\alpha'h$ account
%for the numerically generated dispersion error (or phase error) and the dissipation error (or amplitude error),
%respectively.

To improve the dispersive accuracy for $\alpha'$,
the exact value $\alpha h$ should be very close to $\Re[\alpha' h]$,
where $\Re[\alpha' h]$ denotes the real part of $\alpha' h$.
To achieve the goal of improving solution accuracy,
the positive-value error function $E(\alpha)$
defined below should be very small over the following integration interval for
the modified wavenumber $\alpha h$
\begin{align}
   E(\alpha) = \int_{0}^{\frac{7\pi}{8}}
               \left[ W \cdot \left(\alpha\:h-\Re[\alpha'\:h]\right) \right]^2 d(\alpha h).
     \label{eq:error_DRP}
\end{align}
In Eq.~(\ref{eq:error_DRP}) the weighting function $W$
is chosen to be the denominator of $\left(\alpha\:h-\Re[\alpha'\:h]\right)$. %\cite{bib:Ashcroft190(2003)459-477}.
This choice facilitates us to integrate $E(\alpha)$ exactly.
To make the error function defined in $0\leq\alpha h\leq\frac{7\pi}{8}$
to be positive and minimal,
two extreme conditions given by
\begin{align}
  \frac{\partial E}{\partial c_{2}} = 0,
     \label{eq:error_DRP1}
\end{align}
\begin{align}
   \frac{\partial E}{\partial c_{3}} = 0.
     \label{eq:error_DRP2}
\end{align}
are enforced.
These two constraint equations enforced for maximizing the dispersion accuracy
are used together with the other six algebraic equations derived
from the modified equation analysis to get not only a smaller dissipation error
but also an improved dispersion accuracy.
Note that several integration ranges have been numerically determined so as to find the best one that renders the smallest value of $E$.

The resulting eight introduced unknown coefficients can be determined as
%$a_1 = 0.88825179$,
%$a_3 = 0.04922965$,
%$ b_1 = 0.15007240$,
%$ b_2 = -0.25071279$,
%$ b_3 = -0.01241647$,
%$c_1 = 0.01666172$,
%$c_2 = -1.97080488$ and
%$c_3 = 1.95414316$
$a_1 = 0.8873686$,
$a_3 = 0.0491178$,
$ b_1 = 0.1495320$,
$ b_2 = -0.2507682$,
$ b_3 = -0.0123598$,
$c_1 = 0.0163964$,
$c_2 = -1.9692791$ and
$c_3 = 1.9528828$
from the above reduction of dispersion and dissipation errors.
The upwinding scheme developed theoretically in four stencil points $i-2$, $i-1$,
$i$ and $i+1$ for $\frac{\partial u}{\partial x}$
has the spatial accuracy of order fifth
according to the derived modified equation given below
\begin{align}%\nonumber
\label{eq:order1}
    \frac{\partial u}{\partial x} = \frac{\partial u}{\partial x}|_{exact}
  & +0.0000077381655315119445~{h}^{5}\: \frac{\partial^{6} u}{\partial x^{6}}+ H. O. T.~.
\end{align}
%For $\underline{u}<0$,
%the proposed three-point UCCD scheme can be similarly derived.
It is noted that, unlike our strategy,
Zhou et al. \cite{bib:Zhou227(2007)} chose the coefficient $c_3$ as free parameter
so that the other seven coefficients are expressed as the linear functions of $c_3$ by Taylor's expansion.
Then
these eight coefficients were numerically determined by the standard sequential quadratic programming (SQP) method \cite{bib:Schittkowski1985}
to minimize the error function shown in Eq.~(\ref{eq:error_DRP}). However, this optimization result is
highly sensitive to the initial guess of $c_3$, as pointed by Zhou et al. \cite{bib:Zhou227(2007)}.

Define first the values of $u$ at the half nodal points $i\pm \frac{1}{2}$ as follows:
\begin{align}
\label{eq:uccd11}
\hat{u}_{i+1/2}=\overline{\gamma}_{1} u_{i-1} + \overline{\gamma}_{2} u_{i}
- [(\overline{\alpha}_{1} u_{i-1/2}+\overline{\alpha}_{2} u_{i+3/2})
+ h(\overline{\beta}_{1} u^{'}_{i-1/2}
+ \overline{\beta}_{2} u^{'}_{i+1/2} + \overline{\beta}_{3}u^{'}_{i+3/2} ) ],
\end{align}
and
\begin{align}
\label{eq:uccd22}
\hat{u}_{i-1/2}=\overline{\gamma}_{1} u_{i-2} + \overline{\gamma}_{2} u_{i-1}
- [(\overline{\alpha}_{1} u_{i-3/2}+\overline{\alpha}_{2} u_{i+1/2})
+ h(\overline{\beta}_{1} u^{'}_{i-3/2}
+ \overline{\beta}_{2} u^{'}_{i-1/2} + \overline{\beta}_{3}u^{'}_{i+1/2} ) ].
\end{align}
One can then substitute them into Eq.~(\ref{eq:re-con1}) to get the algebraic equation
for $\frac{\partial u}{\partial x}$ at the node $i$.
Derivation of $\overline{\alpha}_{i}$,
$\overline{\beta}_{i}$ and $\overline{\gamma}_{i}$ is then followed by comparing the coefficients
derived in Eq.~(\ref{eq:CCD-EQ1-1}) for $\frac{\partial u}{\partial x}|_i$.
After a term-by-term comparison of Eq.~(\ref{eq:re-con1}),
we are led to get the coefficients as follows:
%which are
%shown in Eqs.~(\ref{eq:uccd11}) and (\ref{eq:uccd22}).
$\overline{\alpha}_{1}=0.8873686$, $\overline{\alpha}_{2}=0.0491178$,
$\overline{\beta}_{1}=0.1495320 $, $\overline{\beta}_{2}=-0.2507682 $, $\overline{\beta}_{3}=- 0.0123598 $,
$\overline{\gamma}_{1} =-0.0163964 $, $\overline{\gamma}_{2} =1.9528828 $.
In brief,
$\hat{u}_{i+1/2}$ of DRPCCD5 scheme for $c_{i+1/2}\geq0$ is given by
\begin{align}\nonumber
\hat{u}_{i+1/2}^{DRPCCD+}=-0.0163964 u_{i-1} + 1.9528828 f_{i}
- [(0.8873686 \hat{u}_{i-1/2}+0.0491178 \hat{u}_{i+3/2})\\
+ h(0.1495320 \hat{u}^{~'}_{i-1/2}
 -0.2507682 \hat{u}^{~'}_{i+1/2} - 0.0123598 \hat{u}^{~'}_{i+3/2} ) ].
\end{align}
Thus,
the magnitude of the leading error term in the compact interpolation
is less than the corresponding non-compact interpolation on the same order (see Eqs.~(\ref{eq:order2}) and (\ref{eq:order1})).
$\hat{u}_{i+1/2}$ of DRPCCD5 scheme for $c_{i+1/2} <0$  can be similarly derived:
\begin{align}\nonumber
\hat{u}_{i+1/2}^{DRPCCD-}=1.9528828 u_{i-1} - 0.0163964 u_{i}
- [(0.0491178 u_{i-1/2}+0.8873686  u_{i+3/2})\\
+ h(0.0123598 \hat{u}^{~'}_{i-1/2}
 +0.2507682 \hat{u}^{~'}_{i+1/2} - 0.1495320 \hat{u}^{~'}_{i+3/2} ) ].
\end{align}

\subsection{Weighted essentially non-oscillatory (WENO) scheme}
\ \ \ \
Advection equations admit discontinuous solutions.
Weighted essentially non-oscillatory schemes are designed
to achieve the high order of accuracy at smooth regions and switch to lower order interpolation to avoid oscillations near discontinuities.

%Advection equations admit discontinuous solutions.
%Weighted essentially non-oscillatory schemes are designed for problems with piecewise smooth solutions containing discontinuities.
%In other words,
%the WENO schemes use adaptive stenciling to achieve the high order of accuracy at smooth regions of
%the solution while switching to lower order interpolation to avoid oscillations near discontinuities.

\subsubsection{Fifth-order WENO (WENO5) scheme}
\ \ \ \
The form of the interface flux reconstructed by the WENO5 scheme \cite{bib:Jiang126(1996)202-228} reads
\begin{align}
\label{eq:weno5}
\hat{f}_{j+1/2}=\frac{\omega_1}{3}f_{j-2}-\frac16(7\omega_1+\omega_2)f_{j-1}+\frac16(11\omega_1+5\omega_2+2\omega_3)f_j\notag\\
+\frac16(2\omega_2+5\omega_3)f_{j+1}-\frac{\omega_3}{6}f_{j+2}.
\end{align}
In the above equation,
we write
\begin{align}
\label{eq:omega}
\omega_k=\frac{\widetilde{\alpha}_k}{\sum_k\widetilde{\alpha}_k},
~\widetilde{\alpha}_k=\frac{\widetilde{c}_k}{(\widetilde{\beta}_k+\epsilon)^2},~~k=1,2,3.
\end{align}
The optimal weights are $\widetilde{c}_1 = \frac{1}{10}$, $\widetilde{c}_2 = \frac{6}{10}$ and $\widetilde{c}_3 = \frac{3}{10}$.
%$\epsilon=10^{-6}$ is a small number to prevent division by zero.
A very small number ($\epsilon=10^{-6}$) is used to prevent division by zero.
The smoothness indicators $\widetilde{\beta}_k$ are given to detect large discontinuities and automatically switch to the stencil that generates the least oscillatory reconstruction
by
\begin{align}
\label{eq:weno-smoothness-1}
\widetilde{\beta}_1&=\frac{13}{12}{(f_{i-2}-2f_{i-1}+f_i)}^2+\frac{1}{4}{(f_{i-2}-4f_{i-1}+3f_i)}^2, \notag
\\\widetilde{\beta}_2&=\frac{13}{12}{(f_{i-1}-2f_{i}+f_{i+1})}^2+\frac{1}{4}{(f_{i-1}-f_{i+1})}^2,
\\\widetilde{\beta}_3&=\frac{13}{12}{(f_{i}-2f_{i+1}+f_{i+2})}^2+\frac{1}{4}{(3f_{i}-4f_{i+1}+f_{i+2})}^2.\notag
\end{align}
The WENO5 scheme gives fifth-order accurate results in smooth regions of the solution and is non-oscillatory near discontinuities.

%There exist optimal weights $c_k$,$k = 1,...,r$ such that
%if $\omega_k= c_k\forall k$,
%the resulting interpolation is $(2r - 1)-th$ order accurate.
%The WENO limiting process causes the stencil weights to attain their optimal values where the solution is smooth.
%Across or near a discontinuity,
%the weight of the stencil containing the discontinuity approaches zero to yield a non-oscillatory interpolated flux.
%This is achieved by scaling the optimal weights by smoothness indicators of the respective stencils [7],i.e.,

%\begin{align}
%\label{eq:uccd2}
%\alpha_k=\frac{c_k}{(\beta_k+\epsilon)^m}
%\end{align}
%where $\beta_k$ is the smoothness indicator of the $k$-th stencil,
%$\epsilon$is a small number to prevent division by zero
%and $m$ is chosen such that the weights for non-smooth stencils approach zero quickly
%(in the present study,$m$ = 2 is used for all cases).
%To ensure convexity,
%the weights $\alpha_k$ are normalized as

%\begin{align}
%\label{eq:uccd2}
%\omega_k=\frac{\alpha_k}{\sum_k\alpha_k}
%\end{align}
%The resulting interpolation is thus (2$r$-1)-th order accurate in smooth regions of the solution and non-oscillatory near discontinuities.

\subsubsection{Fifth-order compact-reconstruction WENO (CRWENO5) scheme}
\ \ \ \
The drawback of higher order WENO schemes is the increasingly wide stencil when increasing the order of accuracy.
Therefore,
the CRWENO5 has been constructed using three third-order compact interpolations as candidates \cite{bib:Ghosh34(3)(2012)}.
%The smoothness indicators $\widetilde{\beta}_{i}$
%(\ref{eq:weno-smoothness-1})-(\ref{eq:weno-smoothness-1}),
%are used to compute the weights of each stencil,
%resulting in an implicit system given by
The CRWENO5 scheme can be expressed as
%The numerical flux $\hat{f}_{i+1/2}^{CRWENO+}$ for $f^{'}(u)|_{x=x_{i+1/2}}>0$ is built% through the values $\hat{f}_{i-1/2}$, $\hat{f}_{i+1/2}$ and $\hat{f}_{i+3/2}$:
\begin{align}
\label{eq:crweno}
(\frac23\omega_1+\frac13\omega_2)\hat{f}_{i-1/2}+[\frac13\omega_1+\frac23(\omega_2+\omega_3)]\hat{f}_{i+1/2}+\frac13\omega_3\hat{f}_{i+3/2}\notag\\
=\frac{\omega_1}{6}f_{i-1}+\frac{5(\omega_1+\omega_2)+\omega_3}{6}f_{i}+\frac{\omega_2+5\omega_3}{6}f_{i+1}.
\end{align}
Note that $\hat{f}_{i+1/2}$ in Eq. (\ref{eq:crweno})
is the approximation of the left biased numerical flux $\hat{f}^{CRWENO+}_{i+1/2}$ for $f^{'}(u)|_{x=x_{i+1/2}}\geq0$.
%The left hand side is a tridiagonal system and the convexity of the weights ensures,
%that none of the main diagonal elements are zero.
Since the weights $\omega_k$ in Section 3.3.1 are overly dissipative, 
they are determined, as suggested in the literature \cite{bib:Yamaleev228(2009)4248-4272,bib:Borges227(2008)3191-3211},
using $\widetilde{\alpha}_k$ as
\begin{align}
\label{eq:uccd2}
\widetilde{\alpha}_k=\widetilde{c}_k(1+\frac{\tau}{\epsilon+\widetilde{\beta}_k}),~~k=1,2,3.
\end{align}
Here,
the optimal weights are $\widetilde{c}_1=\frac{1}{5}$, $\widetilde{c}_2=\frac{1}{2}$ and $\widetilde{c}_3=\frac{3}{10}$.
$\tau$ is simply defined as the absolute difference between $\beta_{0}$ and $\beta_{2}$.
%Note that $\hat{f}^{CRWENO-}_{i+1/2}$
%is the approximation of the right biased numerical flux $\hat{f}^{CRWENO-}_{i+1/2}$ for $f^{'}(u)|_{x=x_{i+1/2}}<0$ in Eq. (\ref{eq:crweno}).
%Note that the approximation of the right biased numerical $\hat{f}^{CRWENO-}_{i+1/2}$ for $f^{'}(u)|_{x=x_{i+1/2}}<0$ can be similarly derived.
%The factor,
%$\tau$,
%is defined as the absolute difference between the left-most and right-most smoothness indicators by \cite{bib:Borges227(2008)3191-3211}.
%while it is defined as the square of the divided difference of the appropriate order by [40].

%Their higher order compact interpolation is shown to have superior spectral properties as compared to a
%non-compact interpolation of the same order (or even higher) [10, 32].

\subsection{Fifth-order optimized dispersion-relation-preserving combined compact difference scheme (ODRPCCD5)}
\ \ \ \
%Modeling of highly advective transport is difficult,
%even in the superficially simple case of one-dimensional constant-velocity flow.
In this section,
we briefly present the hybrid strategy to couple CCD with WENO schemes proposed by \cite{bib:Ren192(2003)365-386,bib:Zhou227(2007),bib:Suna270(2014)238-254} and
our novel hybrid strategy based on the monotonicity-maintenance criteria. Both strategies are compared with each other in Section 5.1.2.
%We first introduce the hybrid scheme in Section 3.4.1 and then explain our new hybrid strategy in Section 3.4.2.

\subsubsection{Hybrid strategy by \cite{bib:Ren192(2003)365-386,bib:Zhou227(2007),bib:Suna270(2014)238-254}}
\ \ \ \
Follow the hybrid strategy of \cite{bib:Ren192(2003)365-386,bib:Zhou227(2007),bib:Suna270(2014)238-254},
the numerical flux $\hat{f}_{i+1/2}$ can be written as
\begin{align}
\label{eq:HS-1}
\hat{f}_{i+1/2}=\sigma_{i+1/2}\hat{f}^{DRPCCD5}_{i+1/2}+
(1-\sigma_{i+1/2})\hat{f}^{CRWENO5}_{i+1/2}.
\end{align}
In the above,
$\sigma_{i+1/2}$ is the weight function and its detailed formulation can be expressed as
\begin{align}
\label{eq:HS-2}
\sigma_{i+1/2}=\mathrm{min}( 1 , \frac{r_{i+1/2}}{r_{c}}),
\end{align}
where
$r_{c}$ is constant and
$r_{i+1/2}$ is a smoothness indicator, determined as
\begin{align}
\label{eq:HS-3}
r_{i+1/2}=\mathrm{min}(r_{i},r_{i+1}),
\end{align}
with
\begin{align}
\label{eq:HS-4}
r_{i}=\frac{|2\Delta f_{i+1/2}\Delta f_{i-1/2}|+\varepsilon_1}
{(\Delta f_{i+1/2})^2+(\Delta f_{i-1/2})^2+\varepsilon_1},
\end{align}
where $\Delta f_{i+1/2}=f_{i+1}-f_{i}$ and $\epsilon_1=10^{-6}$.
%{\color{red}It is obvious that Eq. (\ref{eq:HS-1}) is reduced to the DRPCCD5 scheme when $\sigma_{i+1/2}=1$
%and to the CRWENO5 scheme when $\sigma_{i+1/2}=0$.
%Note that the $r_{c}$ acts as a threshold value .}

\subsubsection{Present hybrid strategy}
\ \ \ \
We first define the monotonic range in our present hybrid strategy.
The field variable $u(x,t)$ is normalized by
\begin{align}
\label{eq:Mon}
\tilde{u}(x,t) = \frac{u(x,t) -  u^{n}_{i-1}}{u^{n}_{i+1} -  u^{n}_{i-1}}.
\end{align}
When substituting
the node values $u^{n}_{i-1}$ and $u^{n}_{i+1}$ into Eq. (\ref{eq:Mon}),
these values can be normalized as $\tilde{u}^{n}_{i-1}=0$ and $\tilde{u}^{n}_{i+1}=1$,
respectively (see Fig. 1).
As shown in Fig.1, we then establish our hybrid strategy based on monotonicity-maintenance criteria by requiring face values
$\tilde{u}_{i+1/2}$:
\begin{align}
\label{eq:Mono}
\tilde{u}^{n}_{i} \leq \tilde{u}_{i+1/2} \leq 1,
\end{align}
and $\tilde{u}_{i-1/2}$:
\begin{align}
\label{eq:Mono1}
0 \leq \tilde{u}_{i-1/2} \leq \tilde{u}^{n}_{i}.
\end{align}
where $\tilde{u}_{i\pm1/2}$ is calculated by substituting face value $\hat{u}_{i\pm1/2}$ into Eq. (\ref{eq:Mon}).

In addition,
the new $\tilde{u}_{i}$ value must be constrained to maintain monotonicity by the following formulation
\begin{align}
\label{eq:Mono3}
\tilde{u}^{n+1}_{i-1} \leq \tilde{u}^{n+1}_{i}  \leq  \tilde{u}^{n+1}_{i+1}.
\end{align}
We discretize Eq.~(\ref{eq:1dwave}) as
\begin{align}
\label{eq:Mono2}
\tilde{u}^{n+1}_{i} =  \tilde{u}^{n}_{i} - \mathbf{Cr} (\tilde{u}_{i+1/2} - \tilde{u}_{i-1/2} ),
\end{align}
where $\mathbf{Cr}=\frac{c\Delta t}{h}$.
Substituting Eq. (\ref{eq:Mono2}) into left-hand inequality of Eq. (\ref{eq:Mono3})  leads to
\begin{align}
\label{eq:Mono4}
\tilde{u}_{i+1/2} \leq \tilde{u}_{i-1/2} + \frac{1}{\mathbf{Cr}} ( \tilde{u}^{n}_{i} - \tilde{u}^{n+1}_{i-1}).
\end{align}
%In Eq.~(\ref{eq:Mono4}),
%the worst-case condition is given by $\tilde{u}_{i-1/2}=0$ and $\tilde{u}^{n+1}_{i-1}=0$
%because $\tilde{u}_{i-1/2}$ is positive and $\tilde{u}^{n+1}_{i-1}$ is negative.
Since $\tilde{u}_{i-1/2}\geq0$ and $\tilde{u}^{n+1}_{i-1}\leq0$,
the worst-case condition in Eq.~(\ref{eq:Mono4}) is given by $\tilde{u}_{i-1/2}=0$ and $\tilde{u}^{n+1}_{i-1}=0$.
It means that Eq.~(\ref{eq:Mono4}) can be rewritten as
\begin{align}
\label{eq:Mono5}
\tilde{u}_{i+1/2} \leq \frac{\tilde{u}^{n}_{i}}{\mathbf{Cr}}.%~~~ for~0<\tilde{u}^{n}_{i} < 1=0
\end{align}
Thus,
the monotonic range can be determined by Eq. (\ref{eq:Mono}), Eq. (\ref{eq:Mono5}) and $0\leq\tilde{u}^{n}_{i} \leq 1$,
as shown in the shadow region in Fig. 2.
The slope of the Courant-number-dependent boundary (dashed line in Fig. 2), $ \frac{1}{\mathbf{Cr}}$,
changes with $\mathbf{Cr}$.
%For $\tilde{u}^{n}_{i} < 0 $ or $ \tilde{u}^{n}_{i} > 1$,

Once the monotonic range is defined,
we then calculate $\hat{u}^{DRPCCD}_{i+1/2}$ and substitute it into Eq. (\ref{eq:Mon}) to get $\tilde{u}_{i+1/2}$,
and estimate whether  $\tilde{u}_{i+1/2}$ locates in the monotonic range. If yes, set $\hat{u}_{i+1/2}=\hat{u}^{DRPCCD}_{i+1/2}$.
If not, set $\hat{u}_{i+1/2}=\hat{u}^{CRWENO}_{i+1/2}$ or $\hat{u}_{i+1/2}=u^{n}_{i}$.
For clarity, the steps are given as follows:\\\\
$\mathbf{Step~1}$:
if $c \geq 0$,
set $\hat{u}^{DRPCCD}_{i+1/2}=\hat{u}^{DRPCCD+}_{i+1/2}$, $\hat{u}^{CRWENO}_{i+1/2}=\hat{u}^{CRWENO+}_{i+1/2}$
and perform $\mathrm{Step~3}$ to $\mathrm{Step~7}$ according to Fig.~3(a).\\
$\mathbf{Step~2}$:
If $c < 0$,
set $\hat{u}^{DRPCCD}_{i+1/2}=\hat{u}^{DRPCCD-}_{i+1/2}$, $\hat{u}^{CRWENO}_{i+1/2}=\hat{u}^{CRWENO-}_{i+1/2}$
and perform $\mathrm{Step~3}$ to $\mathrm{Step~7}$ according to Fig.~3(b).\\
$\mathbf{Step~3}$:
Compute $\mathbf{B}=u_{D}-u_{U}$;
if $|\mathbf{B}|\leq10^{-8}$,
set $\hat{u}_{i+1/2}=u_{C}$.\\
$\mathbf{Step~4}$:
If $|\mathbf{B}| > 10^{-8}$,
compute $\tilde{u}_{C}=(u_{C}-u_{U})/\mathbf{B}$;
if this is less than 0 or greater than 1,
again set $\hat{u}_{i+1/2}=u_{C}$.\\ %else compute $\tilde{u}_{i+1/2}=(\hat{u}^{DRPCCD}_{i+1/2}-u_{U})/\mathbf{B}$.\\
$\mathbf{Step~5}$:
Compute $\tilde{u}_{i+1/2}=(\hat{u}^{DRPCCD}_{i+1/2}-u_{U})/\mathbf{B}$ and $\hat{u}^{CRWENO}_{i+1/2}$.\\
$\mathbf{Step~6}$:
If $\tilde{u}_{i+1/2}<\tilde{u}_{C}$,
set $\hat{u}_{i+1/2}=\hat{u}^{CRWENO}_{i+1/2}$.\\
$\mathbf{Step~7}$:
If $\tilde{u}_{i+1/2}>\tilde{u}_{C}/ \mathbf{Cr}$,
set  $\tilde{u}_{i+1/2}$=$\tilde{u}_{C}/ \mathbf{Cr}$;
if $\tilde{u}_{i+1/2}>1$,
reset  $\tilde{u}_{i+1/2}=1$.
Construct $\hat{u}_{i+1/2}=\tilde{u}_{i+1/2}\mathbf{B}+{u}_{U}$.\\
$\mathbf{Step~8}$:
Calculate face values $\hat{f}_{i\pm 1/2}=c_{i\pm 1/2}\hat{u}_{i+1/2}$ and update into the next time step according to Eq. (\ref{eq:1dwave2}).

Coupling DRPCCD5 with CRWENO5 using this hybrid strategy leads to the ODRPCCD5 scheme.

\section{Fundamental analysis}
%%%%%%%%%%%%%%%%%%%%%%%%%%%%%%%%%%%%%%%%%%%%%%%%%%%%%%%%%%%%%%%%%%%%%%%%%%%%%%%%%%%%%%%
%%%%%%%%%%%%%%%%%%%%%%%%%%%%%%%%%%%%%%%%%%%%%%%%%%%%%%%%%%%%%%%%%%%%%%%%%%%%%%%%%%%%%%%
\subsection{Dispersion and dissipation errors}
%%%%%%%%%%%%%%%%%%%%%%%%%%%%%%%%%%%%%%%%%%%%%%%%%%%%%%%%%%%%%%%%%%%%%%%%%%%%%%%%%%%%%%%
\ \ \ \
The solution for the model equation
\begin{align}
u_{t}+c~u_{x}=0,
\label{eq:1d-wave}
\end{align}
is given by
\begin{align}
u=\hat{u}_{\alpha}(t)e^{\mathbf{i}\alpha x},
\label{eq:nunerical-u}
\end{align}
where $\mathbf{i}\equiv\sqrt{-1}$ and $\hat{u}_{\alpha}$ is the Fourier mode of the wave number $\alpha$.
Differentiation of the above equation leads to
\begin{align}
\frac{\partial u}{\partial x}|_{exact}=\mathbf{i}\alpha h\frac{\hat{u}_{\alpha}}{h}e^{\mathbf{i}\alpha x}.
\end{align}
The approximated derivative term $\frac{\partial u}{\partial x}$ can be similarly written as
\begin{align}
\frac{\partial u}{\partial x}|_{numerical}=\mathbf{i}\alpha^{'} h\frac{\hat{u}_{\alpha}}{h}e^{\mathbf{i}\alpha x}
=(K_{r}+\mathbf{i}K_{i})\frac{\hat{u}_{\alpha}}{h}e^{\mathbf{i}\alpha x}.
\end{align}
Here,
$K_{r}$ and $K_{i}$, denoting the real and imaginary parts of $\alpha^{'} h$ (cf. Eq.~(\ref{eq:transform3})),
account for the dispersion and dissipation errors, respectively.

Fig.~4 shows the dispersion and dissipation characteristics of fifth-order non-compact finite difference scheme (FD5) \cite{bib:Ghosh34(3)(2012)},
fifth-order compact difference (CD5) scheme \cite{bib:Ghosh34(3)(2012)},
eighth-order optimized compact difference (OCD8) scheme \cite{bib:Kim(1996)} and our proposed DRPCCD5 scheme.
It can be seen that the DRPCCD5 scheme has
a better spectral resolution than the OCD8 scheme.
The dispersion property of the DRPCCD5 scheme is better
than those of the other schemes because of the improved dispersive accuracy.
Furthermore, at frequencies with low dispersion error,
the DRPCCD5 scheme has less dissipation than the other schemes.
%Note that, although the DRPCCD5 scheme is more dissipative at higher frequencies,
%the dispersion errors are large there for all schemes and the higher dissipations help
%in filtering out high-frequency errors.

%The dissipation calculated from the present CCD scheme
%is found to be less accurate than the non-dissipative
%centered-type combined compact difference scheme of Lele.

%%%%%%%%%%%%%%%%%%%%%%%%%%%%%%%%%%%%%%%%%%%%%%%%%%%%%%%%%%%%%%%%%%%%%%%%%%%%%%%%%%%%%%%
\subsection{Assessment of the phase speed anisotropy }
%%%%%%%%%%%%%%%%%%%%%%%%%%%%%%%%%%%%%%%%%%%%%%%%%%%%%%%%%%%%%%%%%%%%%%%%%%%%%%%%%%%%%%%
\ \ \ \
In anisotropic two-dimensional problems,
first-order differencing schemes tend to produce phase space errors \cite{bib:Lele103(1992)16-24,bib:Chu140(1998)370-399}.
To evaluate the phase space error of our DRPCCD5 scheme, we take the following two-dimensional advection equation into consideration
\begin{align}
u_{t}+c_{x}~u_{x}+c_{y}~u_{y}=0.
\label{eq:two-dimension-ad-exact}
\end{align}
Here,
$c_{x}=c~\mathrm{cos}(\theta)$ and $c_{y}=c~\mathrm{sin}(\theta)$ denote the velocity
components along the $x$ and $y$ directions,
respectively.
For a wave propagating at the angle $\theta$ ($\equiv\tan^{-1}(\frac{c_{y}}{c_{x}})$) with respect to the $x$-axis,
the numerical phase speed anisotropy can be derived as follows \cite{bib:Chu140(1998)370-399}
\begin{align}
\label{eq:phase-speed}
\Re(\frac{c^{*}}{c})  = \frac{ cos(\theta)\Re[\alpha^{'}h(\alpha h~cos(\theta))]+
                  ~sin(\theta)\Re[\alpha^{'}h(\alpha h~sin(\theta))]}{\alpha h}.
\end{align}
One can find from Fig.~5
that our proposed scheme reproduces phase speed anisotropies much better
than the sixth-order combined compact difference (CCD6) scheme \cite{bib:Chu140(1998)370-399} at all scaled wavenumbers.

%%%%%%%%%%%%%%%%%%%%%%%%%%%%%%%%%%%%%%%%%%%%%%%%%%%%%%%%%%%%%%%%%%%%%%%%%%%%%%%%%%%%%%%
\subsection{Amplification factor, numerical group velocity and numerical phase velocity}
%%%%%%%%%%%%%%%%%%%%%%%%%%%%%%%%%%%%%%%%%%%%%%%%%%%%%%%%%%%%%%%%%%%%%%%%%%%%%%%%%%%%%%%
\ \ \ \
%{\color{red}In \cite{bib:Sengupta26(2006)151-193},
%properties such as numerical amplification factor,
%numerical group velocity and numerical phase velocity  of the numerical schemes are analyzed by choosing one-dimensional wave equation.
%Similarly,
%the present DRPCCD5 and previous sixth-order combined compact difference (CCD6) \cite{bib:Chu140(1998)370-399} spatial discretization schemes
%are used with fourth-order accuracy time integration Runge-Kutta (RK4) scheme
%to analyze these properties.
The properties, such as amplification factor,  numerical group velocity and numerical phase velocity \cite{bib:Sengupta26(2006)151-193},
of the present DRPCCD5 scheme are analyzed by solving the one-dimensional wave equation,
where the fourth-order accuracy Runge-Kutta (RK4) scheme is applied in time evolution.
The present scheme is compared with the previous sixth-order combined compact difference (CCD6) scheme \cite{bib:Chu140(1998)370-399}.
%For the general numerical solution of Eq. (\ref{eq:1d-wave}),
%we identify it as
%\begin{align}
%\label{eq:TK1}
%u(x_{m},t^{n}) = u^{n}_{m}=\int U(\kappa,t^{n})e^{\mathbf{i}\kappa x_{m}}~d\kappa
%\end{align}
The general numerical solution of Eq. (\ref{eq:1d-wave})
is identified as
\begin{align}
u(x_{m},t^{n}) =\int U(\alpha,t^{n})e^{\mathbf{i}\alpha x_{m}}~d\alpha,
%u(x_{m},t^{n}) =\int A_{0}(\alpha) (G_{r}^{2}+G_{i}^{2})^{\frac{n}{2}} e^{\mathbf{i}(\alpha x_{m}-n\beta)}~d\alpha,
\end{align}
such that the initial solution is given by
\begin{align}
\label{eq:TK1}
u(x_{m},t=0)=\int A_{0}(\alpha) e^{\mathbf{i}\alpha x_{m}}~d\alpha.
\end{align}
Note that the $u(x_{m},t^{n})$ can be obtained by substituting the above initial condition as \cite{bib:Sengupta26(2006)151-193}
\begin{align}
\label{eq:TK2}
%u(x_{m},t^{n}) = u^{n}_{m}=\int U(\kappa,t^{n})e^{\mathbf{i}\kappa x_{m}}~d\kappa.
u(x_{m},t^{n}) =\int A_{0}(\alpha) (G_{r}^{2}+G_{i}^{2})^{\frac{n}{2}} e^{\mathbf{i}(\alpha x_{m}-n\beta)}~d\alpha.
\end{align}
In Eq.(\ref{eq:TK2}),
the numerical amplification factor $G(\alpha)$ is defined as $G(\alpha)=G_{r}+\mathbf{i}G_{\mathbf{i}}=\frac{U(\alpha,t^{n+1})}{U(\alpha,t^{n})}$.
The term $\beta$ is obtained as $\mathrm{tan} \beta = -\frac{G_{\mathbf{i}}}{G_{r}}$.
The numerical group speed and numerical phase  velocity are obtained as
%The numerical group velocity is obtained as \cite{bib:Sengupta26(2006)151-193},
\begin{align}
\label{eq:TK3}
\frac{V_{g}(\alpha)}{c}  = \frac{1}{h~\mathbf{Cr}}\frac{d \beta}{d \alpha},
%V_{g}(\alpha)  = \frac{c}{h~\mathrm{C_{r}}}\frac{d \beta}{d k},
\end{align}
\begin{align}
\label{eq:TK4}
\frac{V_{p}(\alpha)}{c} = \frac{\beta}{\omega\Delta t},
%V_{p}(\alpha) = \frac{c\beta}{\omega\Delta t},
\end{align}
where $\mathbf{Cr}= \frac{c\Delta t}{h}=\frac{\omega\Delta t}{\alpha h}$ denotes the Courant number.

%The numerical properties in $\kappa \Delta x,\omega\Delta t$-plane as
%for the solution of Eq. (\ref{eq:1d-wave}) using the $RK4$ scheme for the time integration and the spatial discretization
%by the present DRPCCD5 and CCD6 \cite{bib:Chu140(1998)370-399} schemes.
%The amplification factor for DRPCCD5 and CCD6 spatial discretization schemes is shown in Fig. 4.
In Figs. 6(a) and (b),
the amplification factors are naturally stable over a large range of $\omega \Delta t$ for both DRPCCD5 and CCD6 schemes.
%In Fig. ?, results
%show that DRPCCD5 scheme performs better in comparison with CCD6 scheme
%since one requires neutral stability ($|G|=1$ ) for a chosen numerical method.
Figs. 6(c) and (d) show the comparison of the variations of $\frac{V_{g}}{c}$ in the $\alpha h-\omega\Delta t$ plane for the two numerical
schemes discussed above.
If one defines the area bounded by the contour lines of $\frac{V_{g}}{c} = 0.95$ and $\frac{V_{g}}{c} = 1.05$
as a DRP region, the DRPCCD5 scheme can resolve the DRP region up to
$\alpha h = 2.5$, while the CCD6 scheme only reaches $\alpha h = 1.68$.
It can be clearly seen that the DRPCCD5 scheme has the better DRP property.
Figs. 6(e) and (f) give the contours of the numerical phase speed.
Similarly, defining the DRP region as bounded by $\frac{V_{p}}{c} = 0.95$ and $\frac{V_{p}}{c} = 1.05$,
one can see that the DRPCCD5 scheme resolves a $10\%$ larger DRP region than the CCD6 scheme.

%It can be seen that the ODRPCCD5 scheme converges faster than the other schemes as the computational grid is refined.

\section{Numerical results}
%The local Lax-Friedrichs (LLF) flux method is used to numerically approximate the flux component for each cell boundary.
%%%%%%%%%%%%%%%%%%%%%%%%%%%%%%%%%%%%%%%%%%%%%%%%%%%%%%%%%%%%%%%%%%%%%%%%%%%%%%%%%%%%%%%
\subsection{One-dimensional problems}
%%%%%%%%%%%%%%%%%%%%%%%%%%%%%%%%%%%%%%%%%%%%%%%%%%%%%%%%%%%%%%%%%%%%%%%%%%%%%%%%%%%%%%%
\ \ \ \
The ODRPCCD5 scheme is tested to solve three linear advection equations and one inviscid Burgers' equation.
%The linear advection equation is an example of a scalar hyperbolic PDE,
%while the inviscid Burgers' equation is an example of a scalar nonlinear hyperbolic PDE.
The $L_{2}$-errors and their corresponding spatial rates of convergence are tested for the linear advection problem\#1.
%And then,
The computational costs are compared using different spatial discretization schemes for the linear advection problems\#2.
Two hybrid strategies descried in section 3.4 are used to solve linear advection problem\#2.
Finally,
we extend ODRPCCD5 scheme to solve the one-dimensional Euler equations of the polytropic gas dynamics.

%%%%%%%%%%%%%%%%%%%%%%%%%%%%%%%%%%%%%%%%%%%%%%%%%%%%%%%%%%%%%%%%%%%%%%%%%%%%%%%%%%%%%%%
\subsubsection{Linear advection problem \#1}
%%%%%%%%%%%%%%%%%%%%%%%%%%%%%%%%%%%%%%%%%%%%%%%%%%%%%%%%%%%%%%%%%%%%%%%%%%%%%%%%%%%%%%%
\ \ \ \
The problem with the smooth initial condition $u(x,0)=\mathrm{sin}(2\pi x)$
for Eq. (\ref{eq:1dwave}) with $c=1$ is solved.
Periodic boundary conditions are applied at two boundaries of the region $0 \leq x \leq 1$.
%To compare the computational efficiency of time integration schemes,
%we solve this problem using DRPCCD5 scheme with two different time integrations, i.e.
%the sixth-order implicit symplectic Runge-Kutta scheme (SRK6)
%and the fourth-order explicit Runge-Kutta scheme (RK4).
To compare the computational efficiency of time evolution,
we solve this problem by using the sixth-order implicit symplectic Runge-Kutta scheme (SRK6) and the fourth-order explicit Runge-Kutta scheme (RK4).
The twin-tridiagonal coefficient matrix for the DRPCCD5 scheme is solved by the computationally effective solver,
including twin-forward elimination and twin-backward substitution techniques,
which is described in \cite{bib:Chu140(1998)370-399}.
All the computational times are obtained using a Core i7, 3.40 GHz computer with 64.0 GB of RAM.
%From figure 1 and Table 1,
%we observe these schemes with different girds gives comparable results and comparable CPU times
%As can be seen in Fig. 7 and Table 1, the numerical results and CPU times using these schemes with different girds
%are slightly different.

Table 1 shows that the SRK6 scheme costs more CPU time than the RK4 scheme when the same spatial scheme and grid are used.
Fig. 7 shows that the computational errors mainly come from the spatial discretization, by comparing RK4/WENO5 and RK4/DRPCCD5.
Therefore, we employ the RK4 scheme for time evolution in the following numerical cases.
The $L_{2}$-errors and their corresponding spatial rates of convergence,
by using DRPCCD5, WENO5, and CRWENO5 schemes, are given in Table 2 with time step $\Delta t = 1\times 10^{-5}$.
It can be seen that all schemes can approximately achieve their theoretical order of accuracy.

%We also assess the computational efficiency in terms of the CPU times for the implicit central CCD
%scheme and the current upwinded dual-compact scheme. In this comparison study, the twin-tridiagonal algebraic equation
%will be solved by the computationally effective solver given in [11] at Re = 100. As can be clearly seen from Table 6 that the
%difference in the computational times is negligibly small

%%%%%%%%%%%%%%%%%%%%%%%%%%%%%%%%%%%%%%%%%%%%%%%%%%%%%%%%%%%%%%%%%%%%%%%%%%%%%%%%%%%%%%%
\subsubsection{Linear advection problem \#2}
%%%%%%%%%%%%%%%%%%%%%%%%%%%%%%%%%%%%%%%%%%%%%%%%%%%%%%%%%%%%%%%%%%%%%%%%%%%%%%%%%%%%%%%
\ \ \ \
%The time integration is performed by means of a three-stage, TVD Runge¨CKutta scheme [54], which is also
%adopted in all other numerical tests in the present paper
The one-dimensional linear equation $u_{t}+ u_{x}=0$ is solved
considering the following initial condition \cite{bib:Abedian185(2014)106-127}:
\begin{align}
u(x,0)=\left\{
\begin{array}{lcl}   %{cc}
\frac{1}{6}(G(x,z-\delta)+G(x,z+\delta)+4G(x,z))    &;&-0.8 \leq x \leq -0.6\\
1                                                   &;&-0.4 \leq x \leq -0.2\\
1-|10(x-0.1)|                                       &;&~0 \leq x \leq 0.2\\
\frac{1}{6}(F(x,a-\delta)+F(x,a+\delta)+4F(x,a))    &;&0.4 \leq x \leq 0.6 \\
0                                                   &;&\mathrm{otherwise}.\\
\end{array}
\right.
\end{align}
where
$G(x,z)=e^{-\beta{(x-z)^2}}$, $F(x,a)={(\mathrm{max}(1-{\alpha^2}{(x-a)^2},0))}^{1/2}$.
The constants are taken as $a=0.5$,\:$z=-0.7$,\:$\delta=0.005$,\:$\alpha=10$,\:and $\beta$=(log 2)/36$\delta^2$.
This initial condition consists of a
discontinuous square wave, an exponential wave, a triangular wave, and a parabolic wave.
Periodic boundary conditions are imposed here.
The time step is chosen as $\Delta t =0.05h$.
Fig.~8 shows the exact waveform and the waveform obtained by the WENO5 and ODRPCCD5 scheme
on a grid with $200$ points at $t=2$ and $t=4$.
Figs.~9 and 10 show the magnified solution for the exponential and square waves at $t=4$.  % exponential
In Fig.~9,
one can see that the ODRPCCD5 show less clipping at the extreme than the WENO5 in the case of the exponential wave.
In Fig.~10,
the ODRPCCD5 scheme is less dissipative than the WENO5 scheme across the discontinuities.
Since the DRPCCD5 scheme is not classified to be a non-oscillatory scheme,
the predicted kinks near the root of square wave is computationally inevitable.
%The results after $one$ cycle over the periodic domain
%are also plotted in Fig. ? with $200\times 200$ grid points.
Comparing the magnitude of errors produced by WENO5, DRPCCD5 and ODRPCCD5
for this test problem shows that ODRPCCD5 performs better.

The computational costs using WENO5, DRPCCD5 and ODRPCCD5 schemes are compared based on different grids, as shown in Table 3.
The ODRPCCD5 scheme needs more CPU time than the other two schemes if the same grid is used because this scheme is hybrid.
However, the spectral properties of the ODRPCCD5 scheme imply that it may apply a coarser grid to achieve the same resolution as the WENO5 scheme at the same order of convergence.
As shown in Fig. 11, the ODRPCCD5 scheme with 600 grids reaches a better resolution than WENO5 scheme with 1600 grids.
Meanwhile, it only needs 2.37$s$ in comparison with 3.9$s$ by WENO5 scheme.

The two hybrid strategies introduced in section 3.4 are used to solve the advection equations, and the numerical results are plotted in Fig. 12
with $400$ grids and $\Delta t = 0.05 h$ at $t=2.0$.
In Fig. 12,
we can see that the solution is not damped when using the previous hybrid strategy by \cite{bib:Ren192(2003)365-386,bib:Zhou227(2007),bib:Suna270(2014)238-254}
when $r_{c}=0.1$.
Therefore, this hybrid strategy needs an appropriate trial parameter ($r_{c}$) to damp the oscillation.
In contrast, our hybrid strategy is based on the monotonicity-maintenance criteria,
which automatically limits oscillations and captures discontinuities, as shown in Fig. 12(b).

%The time integration is performed by means of a three-stage, TVD Runge¨CKutta scheme [54], which is also
%adopted in all other numerical tests in the present paper

%%%%%%%%%%%%%%%%%%%%%%%%%%%%%%%%%%%%%%%%%%%%%%%%%%%%%%%%%%%%%%%%%%%%%%%%%%%%%%%%%%%%%%%
\subsubsection{Linear advection problem \#3}
%%%%%%%%%%%%%%%%%%%%%%%%%%%%%%%%%%%%%%%%%%%%%%%%%%%%%%%%%%%%%%%%%%%%%%%%%%%%%%%%%%%%%%%
\ \ \ \
We solve the linear equation $u_{t}+ u_{x}=0$, $-1\leq x\leq1$,
with periodic boundary condition \cite{bib:Harten24(2)(1987)}.
The initial condition reads
%\begin{align}
%\label{eq:hea}
%g(x)=-(\sqrt{3}/2+9/2+2\pi/3)(x+1) \notag\\
%+\left\{
%\begin{array}{lcl}   %{cc}
%2\cos(3\pi{x^2}/2)-\sqrt{3}                 &;&-1< x < -\frac{1}{3}\\
%3/2+3\cos{2\pi x}                           &;&-\frac{1}{3}< x < 0\\
%(28+4\pi+\cos{3\pi x})/3+6\pi \cos{3\pi x}  &;&\frac{1}{3} < x < 1 \\
%\end{array}
%\right.
%\end{align}
\begin{align}
\label{eq:hea}
u(x,t=0)=
\left\{
\begin{array}{lcl}   %{cc}
-x~\mathrm{sin}(\frac{3\pi x^{2}}{2})                &;&-1< x < -\frac{1}{3}\\
|\mathrm{sin}(2\pi x)|                           &;&-\frac{1}{3}< x < 0\\
2x-1-\frac{1}{6}\mathrm{sin}(3\pi x)  &;&\frac{1}{3} < x < 1 \\
\end{array}
\right.
\end{align}
%This function is the integral of the function used by Harten et al. [5] modulo some
%linear function chosen so that $g(-1)=g(1)$.
%We shift the function to position the singular points of g inside the interval.
The predicted results in the domain with $200$ grid points are plotted in Fig.~13 at $t=20$.
It can be seen that ODRPCCD5 scheme performs better than the WENO5 scheme.

%%%%%%%%%%%%%%%%%%%%%%%%%%%%%%%%%%%%%%%%%%%%%%%%%%%%%%%%%%%%%%%%%%%%%%%%%%%%%%%%%%%%%%%
\subsubsection{Non-linear advection problem}
%%%%%%%%%%%%%%%%%%%%%%%%%%%%%%%%%%%%%%%%%%%%%%%%%%%%%%%%%%%%%%%%%%%%%%%%%%%%%%%%%%%%%%%
\ \ \ \
We solve the Burgers' equation $u_{t}+ (0.5u^{2})_{x}=0$,\;$-1\leq x\leq1$,
with periodic boundary condition.
The initial condition is $u(x,0)=2+ \mathrm{sin}(\pi (x+1))$. %\cite{bib:Harten24(2)(1987)}.
The solution to Burgers' equation is smooth for $t<\frac{1}{\pi}$ and it develops shocks for $t=\frac{1}{\pi}$ .
The results obtained at $t=0.3$ (before shock) and $t=0.35$ (after shock)
are plotted in Fig.~14 in the domain with $200$ grid points.
The time step is chosen as $\Delta t =0.1h$ in this computation.
%We observe that DRPCCD5/CRWENO5 reduces smearing near high gradient region compared with WENO5 schemes.
%In addition,
We observe that ODRPCCD5 gives better results than the DRPCCD5 scheme at $t=0.35$.

%our computed solution of UCCD5/CRWENO5 is in good agreement with the WENO scheme as shown in Fig. ??

%The WENO scheme and the CRWENO5 scheme smooth out the corners in the graph of $g$ as the time increases.

%%%%%%%%%%%%%%%%%%%%%%%%%%%%%%%%%%%%%%%%%%%%%%%%%%%%%%%%%%%%%%%%%%%%%%%%%%%%%%%%%%%%%%%
\subsubsection{The Shu-Osher problem}
%%%%%%%%%%%%%%%%%%%%%%%%%%%%%%%%%%%%%%%%%%%%%%%%%%%%%%%%%%%%%%%%%%%%%%%%%%%%%%%%%%%%%%%
\ \ \ \
In this case, we solve the one-dimensional Euler equations of gas dynamics \cite{bib:Suna270(2014)238-254}
%\cite{bib:Suna270(2014)238-254,bib:Abedian185(2014)106-127}
\begin{align}
\frac{\partial}{\partial t}
{\begin{pmatrix}
\rho\\
\rho{q}\\
E
\end{pmatrix}}+
\frac{\partial}{\partial x}
{\begin{pmatrix}
\rho{q}\\
\rho{q}^2+p\\
q(E+p)
\end{pmatrix}}=0,\\
p=(\gamma-1)(E-\frac{1}{2}{\rho}q^2),~\gamma=1.4.
\end{align}
where $\rho$, $q$, $p$ and $E$ are the density, velocity, pressure and total
energy of the conserved fluid, respectively.
The initial conditions are
\begin{align}
(\rho,u,p)=\left\{
\begin{array}{lcl}
(3.857143,2.629369,10.3333)  &,& if~~x\leq1\\
(1+0.2\sin(5x),0,1)          &,& otherwise\\
\end{array}
\right.
\end{align}
This test case leads to very strong shock waves and
is employed to validate the shock-capturing capability of the proposed ODRPCCD5 scheme.
Reflective boundary conditions are applied at both $x=0$ and $x=10$.
Since the exact solution for this problem is not available,
the solution computed in $10000$ grids is considered as the exact solution.
Fig.~15 shows waveforms at $t=0.45$, $t=0.9$, $t=1.35$ and $t=1.8$  (grid spacing $h=\frac{1}{40}$, time step $\Delta t =0.05h$).
It can be seen that the shock-waves are well reproduced by our proposed ODRPCCD5 scheme.
\subsection{Two-dimensional problems}
%%%%%%%%%%%%%%%%%%%%%%%%%%%%%%%%%%%%%%%%%%%%%%%%%%%%%%%%%%%%%%%%%%%%%%%%%%%%%%%%%%%%%%%
\ \ \ \
In this subsection,
we illustrate the capacity of the ODRPCCD5 scheme
through two-dimensional numerical simulations.
%%%%%%%%%%%%%%%%%%%%%%%%%%%%%%%%%%%%%%%%%%%%%%%%%%%%%%%%%%%%%%%%%%%%%%%%%%%%%%%%%%%%%%%
\subsubsection{Vortex flow problem}
%%%%%%%%%%%%%%%%%%%%%%%%%%%%%%%%%%%%%%%%%%%%%%%%%%%%%%%%%%%%%%%%%%%%%%%%%%%%%%%%%%%%%%%
\ \ \ \
The equation $\phi_{t}+(u\phi)_{x}+(v\phi)_{y}=0$ is solved using an initial
circle shape in a square of unit length,
within which the vortex flow field $(u,v)$ is given by \cite{bib:Olsson2005} %\cite{bib:Wang35(2009)}
%\begin{align}
%   \label{eq:vortex-velocity}
%u=\mathrm{sin}^{2}(\pi x)\mathrm{sin}(2\pi y)\mathrm{cos}(\frac{\pi t}{T}),\\
%v=\mathrm{sin}^{2}(\pi y)\mathrm{sin}(2\pi x)\mathrm{cos}(\frac{\pi t}{T}).
%\end{align}
\begin{align}
   \label{eq:vortex-velocity}
u=-\mathrm{sin}^{2}(\pi x)\mathrm{sin}(2\pi y),\\
v=\mathrm{sin}^{2}(\pi y)\mathrm{sin}(2\pi x).
\end{align}
The radius of the circle is $0.15$ located at the center $(0.5,0.75)$.
At $t=T$ the flow field was reversed,
so that the exact solution at $t=2T$ should coincide with the initial condition.
This problem has been known to be computationally challenging since its solution
is stretched and torn by the vortex flow where a
very thin filament having a scale of single mesh size can be generated.
%This test problem is solved with $\Delta t= \frac{1}{1000}$.

Computations were performed for $T=2.5$ and $\Delta t= \frac{1}{1000}$.
The predicted results of WENO5 and ODRPCCD5 are compared for the calculation of $\phi=0$.
The results obtained in $100 \times 100$ grids at $t=1.5,~2.5,~4, 5$ are plotted in Fig.~16.
It is clear that the solution computed using the ODRPCCD5 scheme is maintained within a thin and
elongated filament on the scale of one grid spacing.
On the contrary,
the WENO5 scheme results in a considerable reduction of the area at the head and tail of the filament.
In Fig.~16(d), one can see that the solution computed using our proposed scheme returns to its initial state.
In Fig. 17,
the ODRPCCD5 scheme using a $100\times100$ mesh can reach the same resolution at $t=2.5$ as the WENO5 scheme using a $200\times200$ mesh.
Hence, the ODRPCCD5 needs less CPU time (8.30$s$) than the WENO5 scheme (17.32$s$).

%%%%%%%%%%%%%%%%%%%%%%%%%%%%%%%%%%%%%%%%%%%%%%%%%%%%%%%%%%%%%%%%%%%%%%%%%%%%%%%%%%%%%%%
\subsubsection{Zalesak's problem}
%%%%%%%%%%%%%%%%%%%%%%%%%%%%%%%%%%%%%%%%%%%%%%%%%%%%%%%%%%%%%%%%%%%%%%%%%%%%%%%%%%%%%%%
\ \ \ \
The Zalesak's problem \cite{bib:Zalesak1979,bib:Ii231(2012)2328-2358}
is one of the best known benchmark cases for testing the developed advection scheme.
The slotted disk has a radius of 15 and a slot width of 5.
It is initially located at (50,75) in the domain of size (100,100).
The prescribed velocity field is given as
\begin{align}
   \label{eq:zalesak-velocity}
(u=\frac{\pi(50-y)}{314}, v=\frac{\pi(x-50)}{314}).
\end{align}
The results predicted for $100\times100$ grid points at $t=50\pi$, $t=100\pi$, $t=150\pi$ and $t=200\pi$
are plotted in Fig.~18(a).
The results are also plotted in Fig.~18(b) in the domain with $200\times 200$ grid points.
The solution computed with the proposed scheme is in good agreement with the exact (or initial) solution as shown in Fig.~18(b).

\section{Concluding remarks}
\ \ \ \
In this paper,
a fifth-order dispersion-relation-preserving combined compact difference (DRPCCD5) scheme has been proposed, which shows increased dispersion accuracy and improved
dispersion-relation-preserving properties compared to the CCD6 \cite{bib:Chu140(1998)370-399} scheme.
To make discontinuity capturing possible and handle large gradients, an optimized DRPCCD5 scheme (ODRPCCD5), which couples the DRPCCD5 and CRWENO5 schemes, is constructed
using a novel hybrid strategy based on the monotonicity-maintenance criteria.
%To verify the scheme, the computed $L_2$-error norms and their resulting spatial rates of convergence are calculated for the sin wave problem.
%The numerical solutions of linear problem\#2, including a discontinuous square wave, an exponential wave, a triangular wave, and a parabolic wave profile,
%show that our ODPRCCD5 scheme performs very well and is much faster than the previous WENO5 scheme at the same accuracy.
The numerical solutions of linear problems show that our ODPRCCD5 scheme performs very well and is much faster than the previous WENO5 scheme at the same accuracy.
In addition, the ODPRCCD5 scheme produces non-oscillatory solutions of the Euler equations in domains with discontinuities,
%can handle sharp resolutions and remains essentially non-oscillatory near discontinuities when solving the two-dimensional vortex flow and Zalesak's problems.
and it can handle sharp resolutions when solving the two-dimensional vortex flow and Zalesak's problems.
%We show that the present hybrid DRPCCD5/CRWENO5 scheme can handle
%the essentially non-oscillatory near discontinuities for the Shu-Osher problem.
%the proposed scheme is extended to the Euler equations of fluid dynamics.
%It is necessary to yield non-oscillatory results by using proposed scheme for Euler flows with strong discontinuities.
%We show that the present hybrid DRPCCD5/CRWENO5 scheme can handle
%the essentially non-oscillatory near discontinuities.
We plan to apply our algorithm to solve the three-dimensional Navier-Stokes equations for the simulation of two-phase flows in future studies.

%Also,
%we straightforwardly extended this procedure to two-dimensional cases.
%We show that the present hybrid DRPCCD5/CRWENO5 scheme can handle
%the sharp solution regions and remains essentially non-oscillatory near discontinuities.

%%%%%%%%%%%%%%%%%%%%%%%%%%%%%%%%%%%%%%%%%%%%%%%%%%%%%%%%%%%%%%%%%%%%%%%%%%%%%%%%%%%%%%%%%%%
\section*{Acknowledgement}
%%%%%%%%%%%%%%%%%%%%%%%%%%%%%%%%%%%%%%%%%%%%%%%%%%%%%%%%%%%%%%%%%%%%%%%%%%%%%%%%%%%%%%%%%%%
This study was partially supported by the Natural Science Foundation of China (41376095),
Zhejiang University Ocean Sciences Seed Grant (2012HY012B),
and Fundamental Research Funds for the Central Universities (2014QNA4030).
%NSC-97-2221-E-002-250-MY3 and CQSE 97R0066-69.
%Partial support from the National Center for Theoretical Sciences is also acknowledged.

%%%%%%%%%%%%%%%%%%%%%%%%%%%%%%%%%%%%%%%%%%%%%%%%%%%%%%%%%%%%%%%%%%%%%%%%%%%%%%%%%%%%%%%%%%%%%%%%%%%%%%%%%%%%

\newpage

%%%%%%%%%%%%%%%%%%%%%%%%%%%%%%%%%%%%%%%%%%%%%%%%%%%%%%%%%%%%%%%%%%%%%%%%%%%%%%%%%%%%%%%%%%%%%%%%%%%%%%%%%%%%

\begin{table}
\centering
\begin{tabular}{lccccccc}\toprule
Scheme  &grids  &&& \multicolumn{3}{c}{CPU times $(s)$}
\\ \hline
SRK6/DRPCCD5 &  40  &   &&&  22.93  \\
             &  80  &   &&&  42.13   \\
             &  160  &   &&&  79.03  \\
             &  320  &   &&&  145.29   \\ \bottomrule
RK4/DRPCCD5  &  40  &   &&&   8.90 \\
             &  80  &   &&&  13.11   \\
             &  160  &   &&&  22.40   \\
             &  320  &   &&&  41,46   \\ \bottomrule
RK4/WENO5    &  40  &   &&&  5.83  \\
             &  80  &   &&&  6.95   \\
             &  160  &   &&&  9.68   \\
             &  320  &   &&&  15.39   \\ \bottomrule
%DRPCCD5/CRWENO5 &  20$\times$20  &   8.065$\times 10^{-6}$  &&&    \\
%                &  40$\times$40  &   1.269$\times 10^{-7}$  &&&  5.989   \\
%                &  60$\times$60  &   1.820$\times 10^{-8}$  &&&  4.789   \\
%                &  80$\times$80  &   7.629$\times 10^{-9}$  &&&  3.023   \\ \bottomrule
\end{tabular}
\caption{Comparisons of the computational costs for the different schemes at $t=1000$ with $\Delta t= 0.001$.
This problem is described in section 5.1.1.}
\end{table}

\begin{table}
\centering
\begin{tabular}{lccccccc}\toprule
Scheme  &grids  & {$L_{2}$~error norms }  &&
\multicolumn{3}{c}{rates of convergence}
\\ \hline
WENO5 &  20  &   3.724$\times 10^{-4}$  &&&    \\
      &  40  &   8.297$\times 10^{-6}$  &&&  5.488   \\
      &  60  &   8.832$\times 10^{-7}$  &&&  5.524   \\
      &  80  &   1.835$\times 10^{-7}$  &&&  5.461   \\ \bottomrule
CRWENO5 &  20  &   8.056$\times 10^{-6}$  &&&    \\
        &  40  &   1.198$\times 10^{-7}$  &&&  6.071   \\
        &  60  &   1.229$\times 10^{-8}$  &&&  5.615   \\
        &  80  &   2.501$\times 10^{-9}$  &&&  5.534   \\ \bottomrule
DRPCCD5 &  20  &   1.207$\times 10^{-6}$  &&&    \\
        &  40  &   2.597$\times 10^{-8}$  &&&  5.539   \\
        &  60  &   2.783$\times 10^{-9}$  &&&  5.508   \\
        &  80  &   5.750$\times 10^{-10}$  &&&  5.482   \\ \bottomrule
%DRPCCD5/CRWENO5 &  20$\times$20  &   8.065$\times 10^{-6}$  &&&    \\
%                &  40$\times$40  &   1.269$\times 10^{-7}$  &&&  5.989   \\
%                &  60$\times$60  &   1.820$\times 10^{-8}$  &&&  4.789   \\
%                &  80$\times$80  &   7.629$\times 10^{-9}$  &&&  3.023   \\ \bottomrule
\end{tabular}
\caption{The predicted $L_{2}$-error norms and the corresponding spatial rates of convergence
for the solutions predicted with SRK6 scheme at $t=1$ in a domain containing four chosen meshes.
This problem is described in section 5.1.1.}
\end{table}

\begin{table}
\centering
\begin{tabular}{lccccccc}\toprule
Scheme  &grids  &&& \multicolumn{3}{c}{CPU times $(s)$}
\\ \hline
WENO5        &  200  &    &&&  0.078  \\
             &  400  &    &&&  0.32   \\
             &  600  &   &&&  0.59  \\
             &  800  &   &&&  1.04  \\
             &  1600  &   &&&  3.91   \\ \bottomrule
DRPCCD5      &  200  &    &&&   0.20 \\
             &  400  &    &&&  0.79   \\
             &  600  &   &&&  1.68   \\
             &  800  &   &&&  2.99   \\
             &  1600  &   &&&  11.43   \\ \bottomrule
ODRPCCD5     &  200  &    &&&  0.26  \\
             &  400  &    &&&  1.21   \\
             &  600  &   &&&  2.37   \\
             &  800  &   &&&  4.99   \\
             &  1600  &   &&&  19.20   \\ \bottomrule
%DRPCCD5/CRWENO5 &  20$\times$20  &   8.065$\times 10^{-6}$  &&&    \\
%                &  40$\times$40  &   1.269$\times 10^{-7}$  &&&  5.989   \\
%                &  60$\times$60  &   1.820$\times 10^{-8}$  &&&  4.789   \\
%                &  80$\times$80  &   7.629$\times 10^{-9}$  &&&  3.023   \\ \bottomrule
\end{tabular}
\caption{Comparisons of the computational costs for the different schemes at $t=4$ with Courant number $0.05$.
This problem is described in section 5.1.2.}
\end{table}

\begin{figure}[!hbp]
\vskip12pt
\centering
\mbox{\subfigure[]{\includegraphics[width=0.6\textwidth]{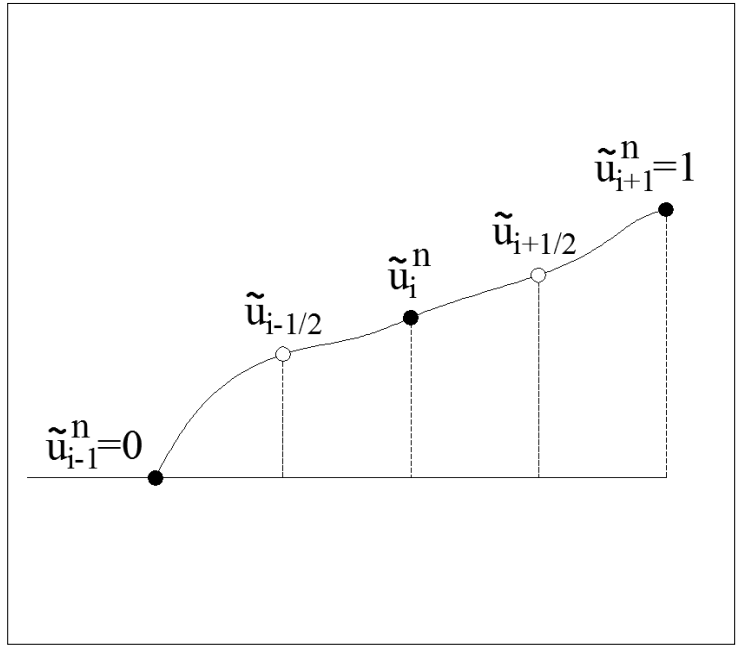}}}
\caption{Location of normalized node and face values for the monotonic behavior.}
\end{figure}

\begin{figure}[!hbp]
\vskip12pt
\centering
\mbox{\subfigure[]{\includegraphics[width=0.6\textwidth]{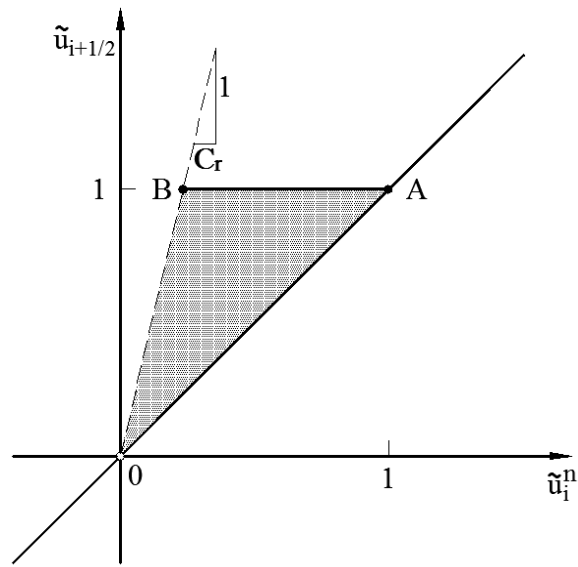}}}
\caption{Monotonic range and normalized variable values.
The dashed line is a Courant-number-dependent slope of $\frac{1}{\mathrm{Cr}}$.}
\end{figure}

\begin{figure}[!hbp]
\vskip12pt
\centering
\mbox{\subfigure[]{\includegraphics[width=0.6\textwidth]{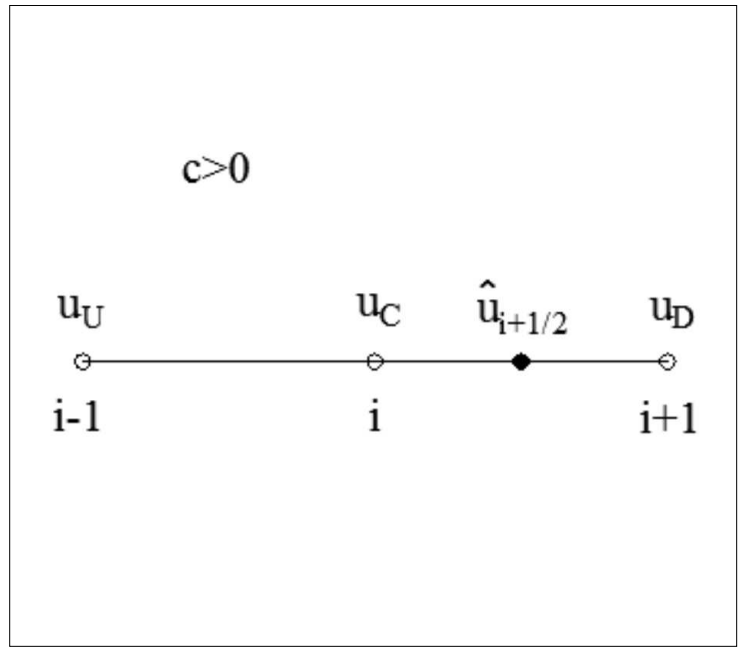}}}
\mbox{\subfigure[]{\includegraphics[width=0.6\textwidth]{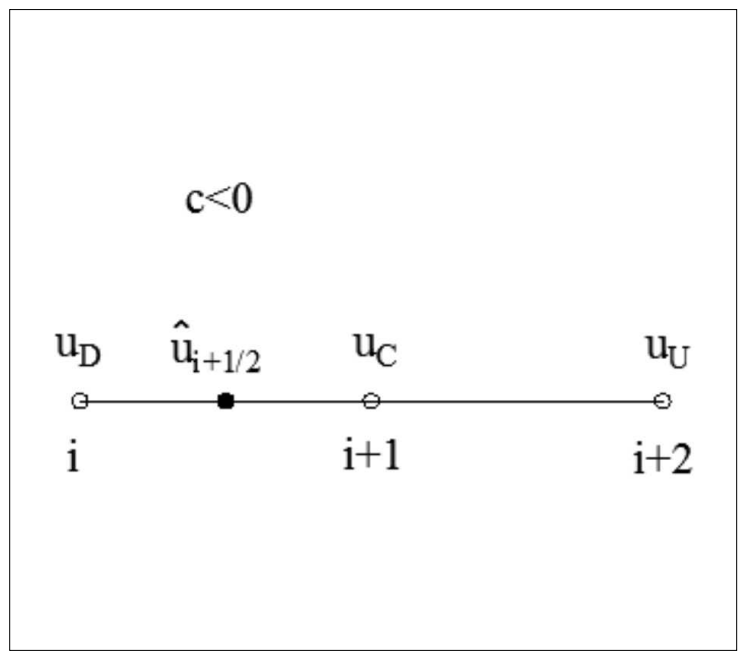}}}
\caption{Definition of upstream (U), downstream (D) and central (C) node-values.
(a) $c>0$;
(b) $c<0$. }
\end{figure}

\begin{figure}[!hbp]
\vskip12pt
\centering
\mbox{\subfigure[]{\includegraphics[width=0.6\textwidth]{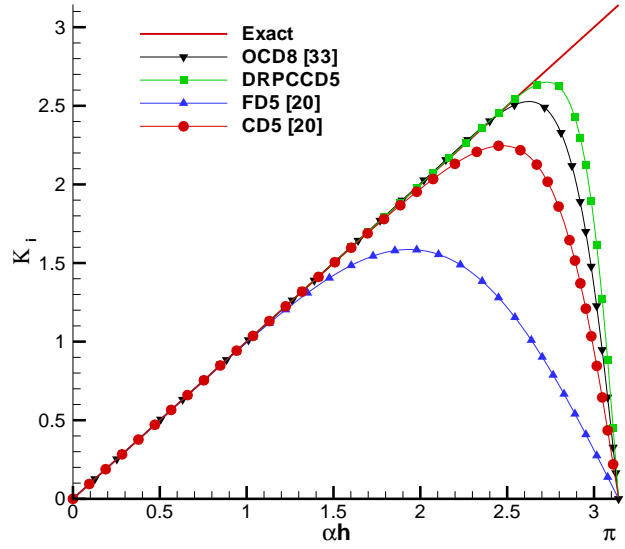}}}
\mbox{\subfigure[]{\includegraphics[width=0.6\textwidth]{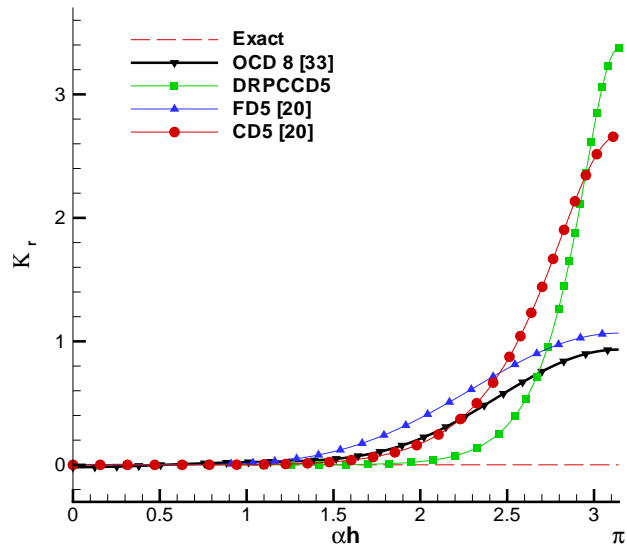}}}
\caption{Comparison of $K_{i}(\alpha h)$ and $K_{r}(\alpha h)$
amongst the proposed DRPCCD5 scheme,
WENO5 scheme \cite{bib:Jiang126(1996)202-228},
CD5 scheme \cite{bib:Ghosh34(3)(2012)},
and OCD8 scheme \cite{bib:Kim(1996)}.
(a) $K_{i}$;
(b) $K_{r}$. }
\end{figure}

%%%%%%%%%%%%%%%%%%%%%%%%%%%%%%%%%%%%%%%%%%%%%%%%%%%%%%%%%%%%%%%%%%%%%%%%%%%%%%%%%%%%%%%%%%%%%%%%%%%%%%%%%%%
\begin{figure}[!hbp]
\vskip12pt
\centering
\mbox{\subfigure[]{\includegraphics[width=0.6\textwidth]{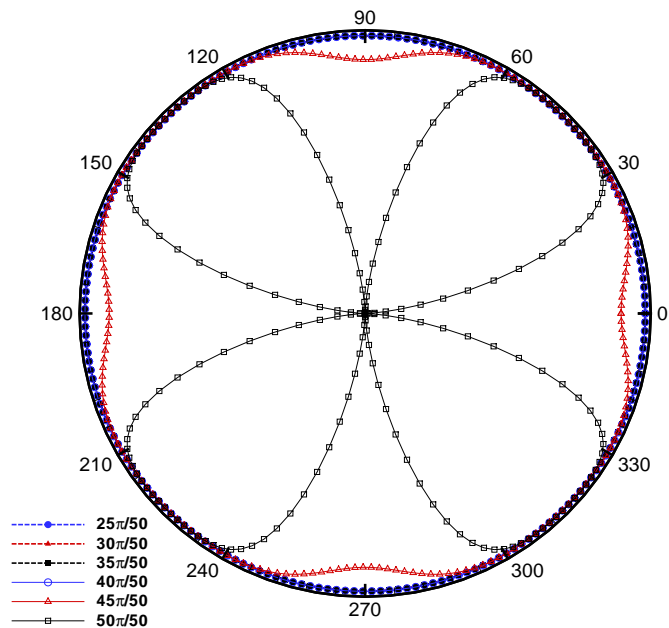}}}
\mbox{\subfigure[]{\includegraphics[width=0.6\textwidth]{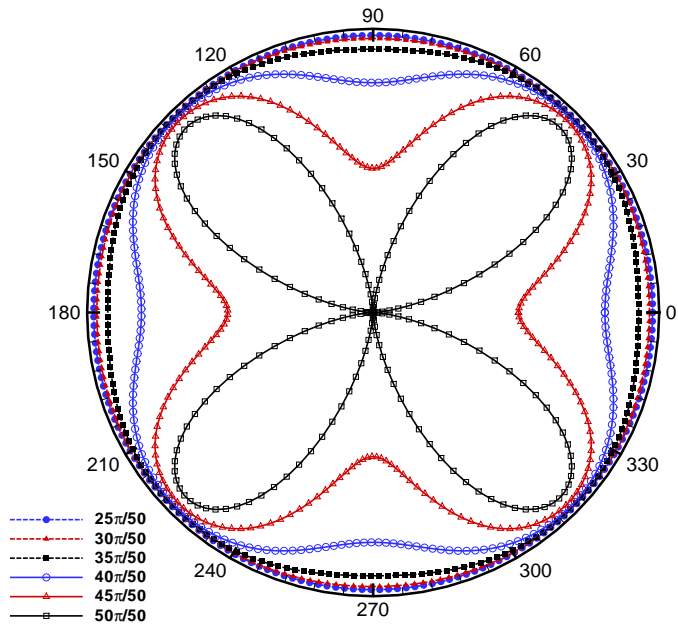}}}
\caption{Comparison of the predicted phase speed anisotropy,
which is plotted against $\theta$, for the proposed DRPCCD5 scheme and the CCD6 scheme of Chu and Fan \cite{bib:Chu140(1998)370-399}.
(a) DRPCCD5 scheme;
(b) CCD6 scheme \cite{bib:Chu140(1998)370-399}.}
\end{figure}

\begin{figure}[!hbp]
\vskip12pt
\centering
\mbox{\subfigure[]{\includegraphics[width=0.41\textwidth]{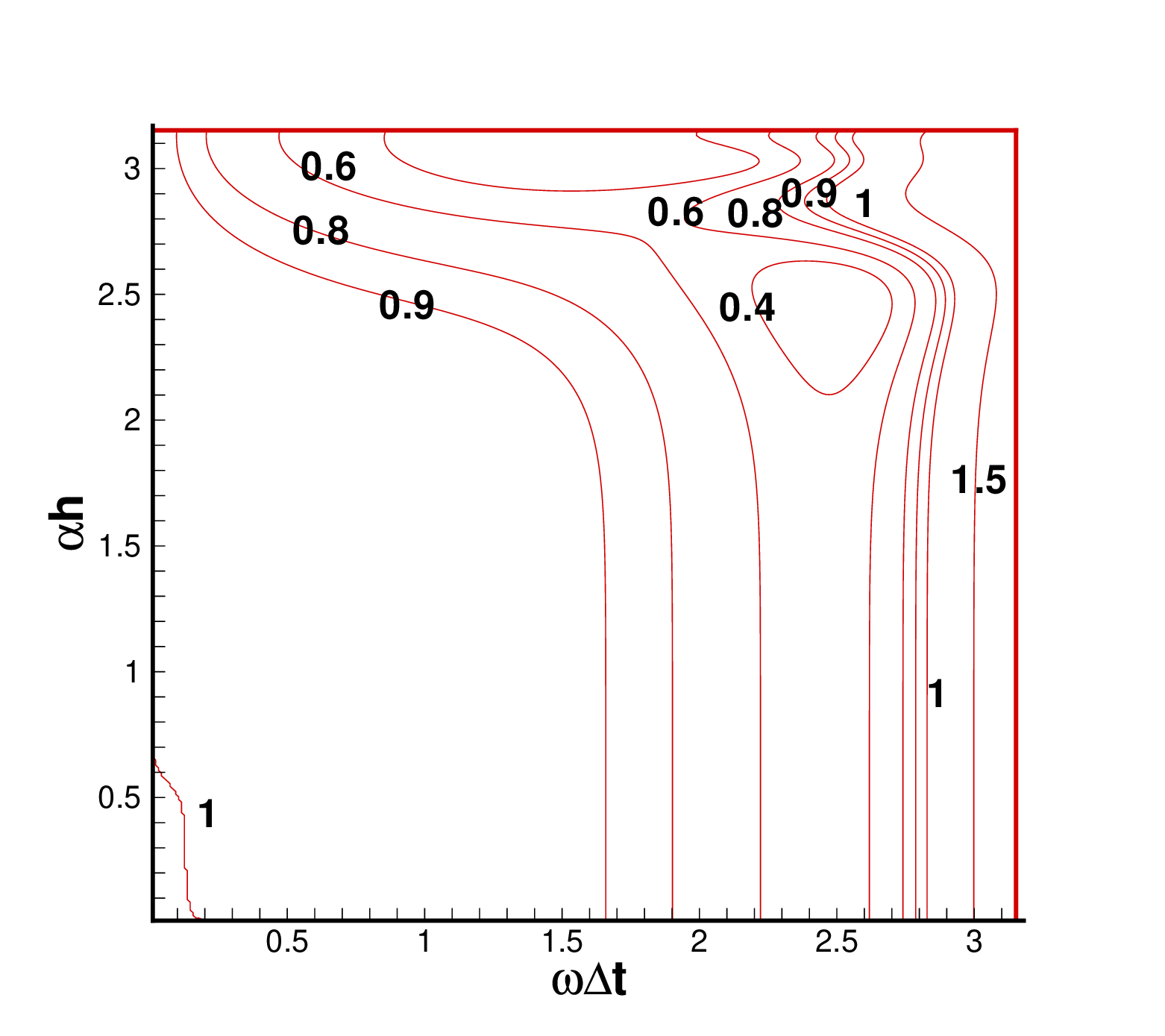}}
      \subfigure[]{\includegraphics[width=0.41\textwidth]{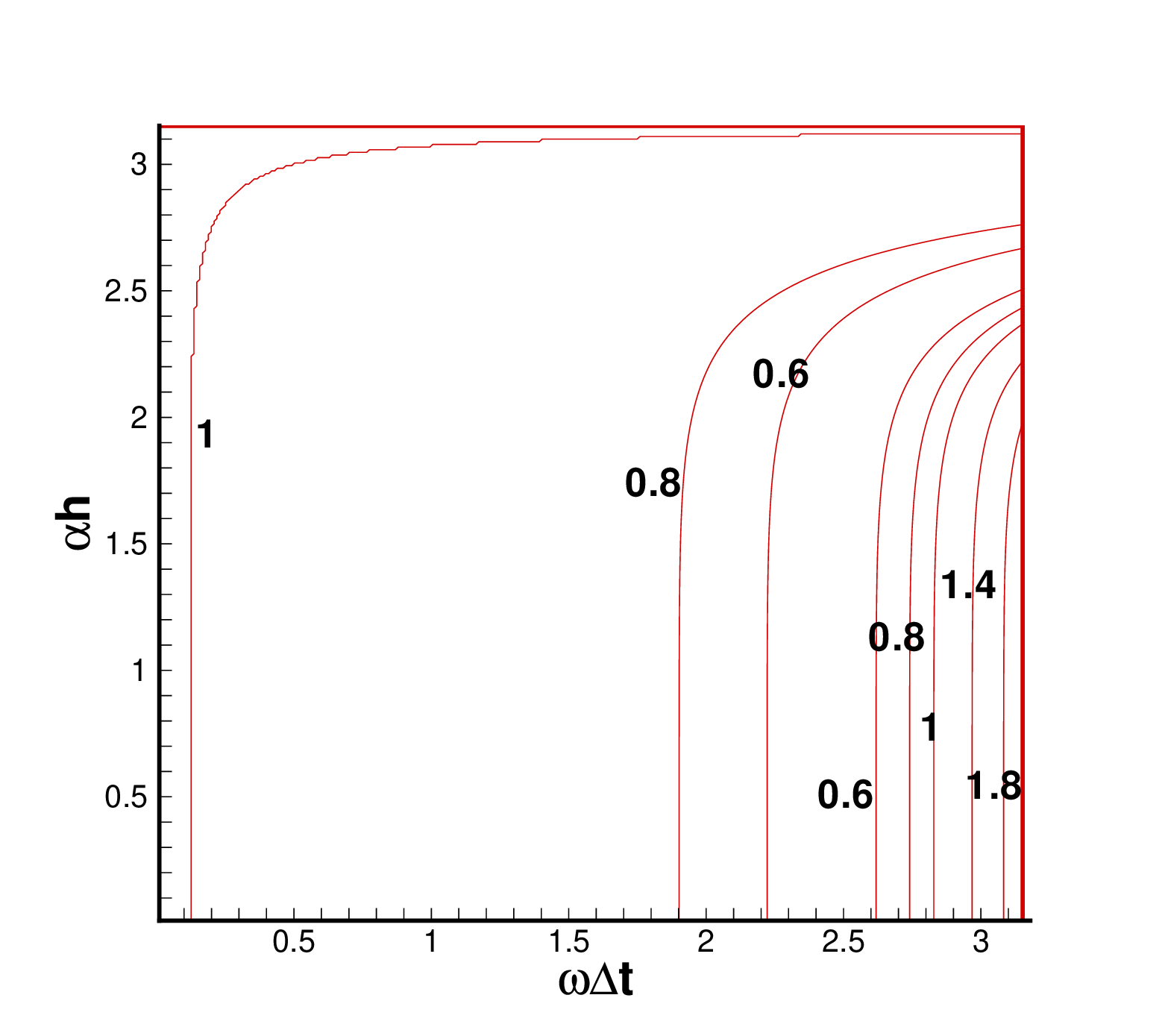}} }
\mbox{\subfigure[]{\includegraphics[width=0.41\textwidth]{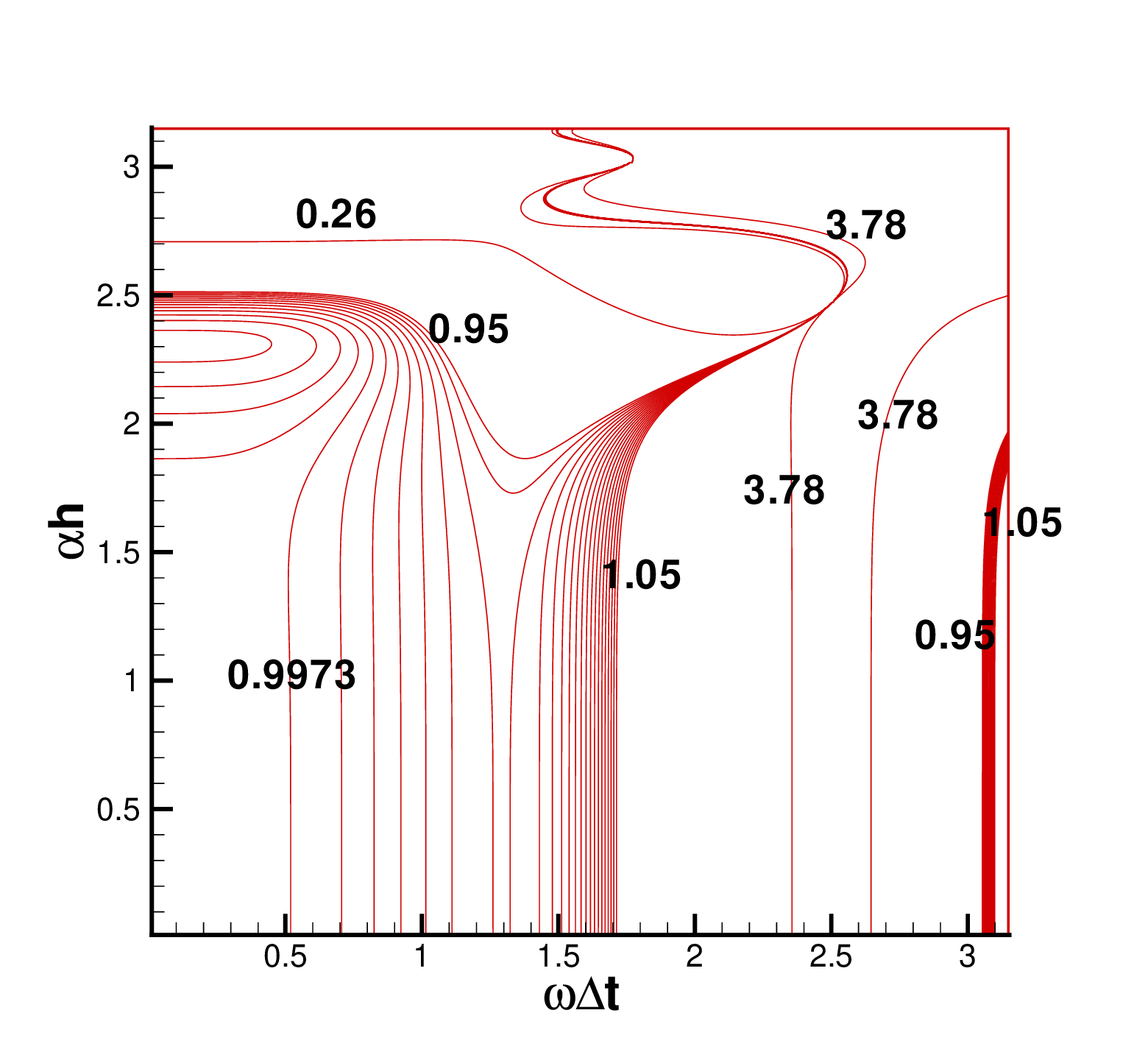}}
      \subfigure[]{\includegraphics[width=0.41\textwidth]{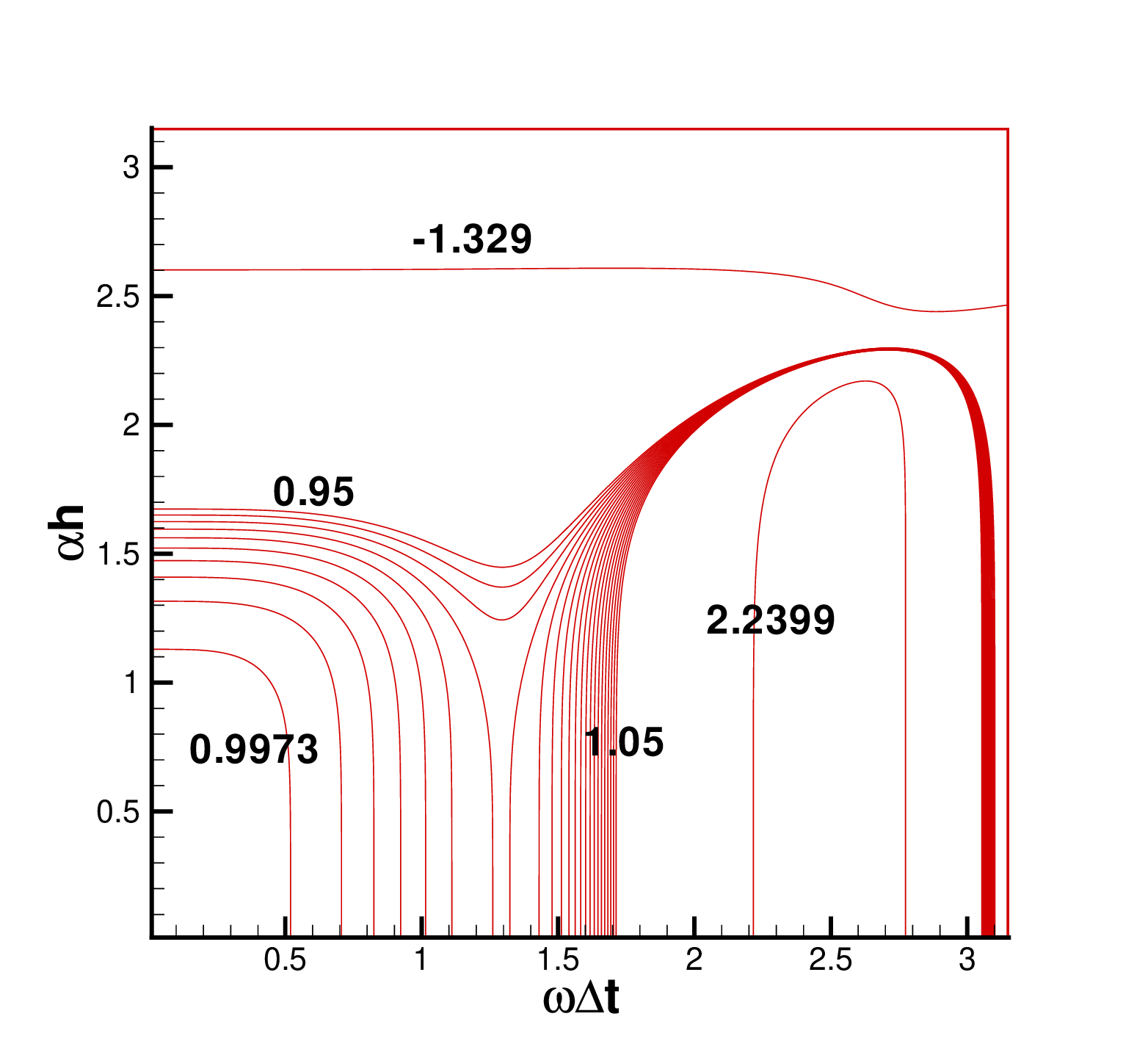}} }
\mbox{\subfigure[]{\includegraphics[width=0.41\textwidth]{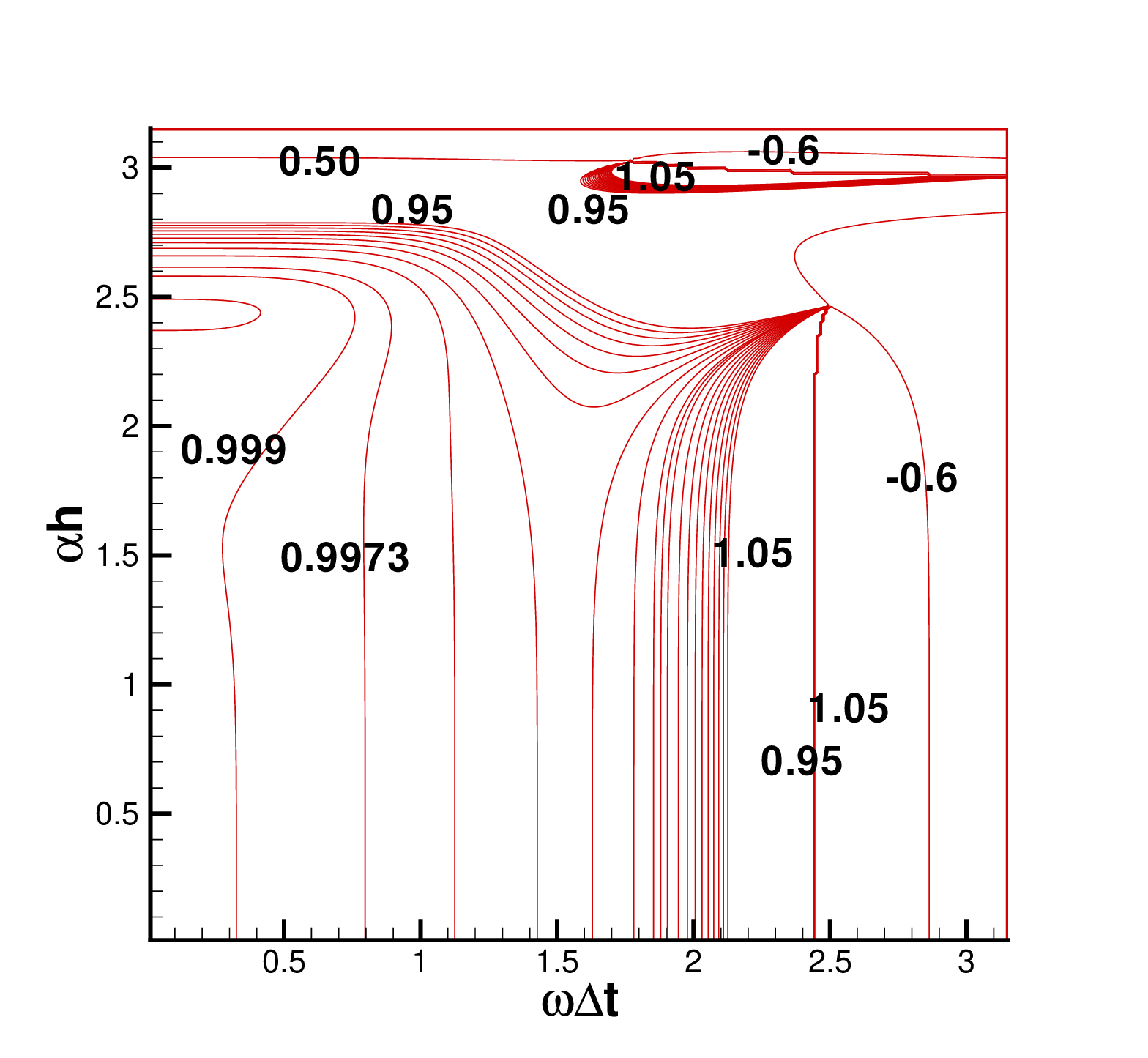}}
      \subfigure[]{\includegraphics[width=0.41\textwidth]{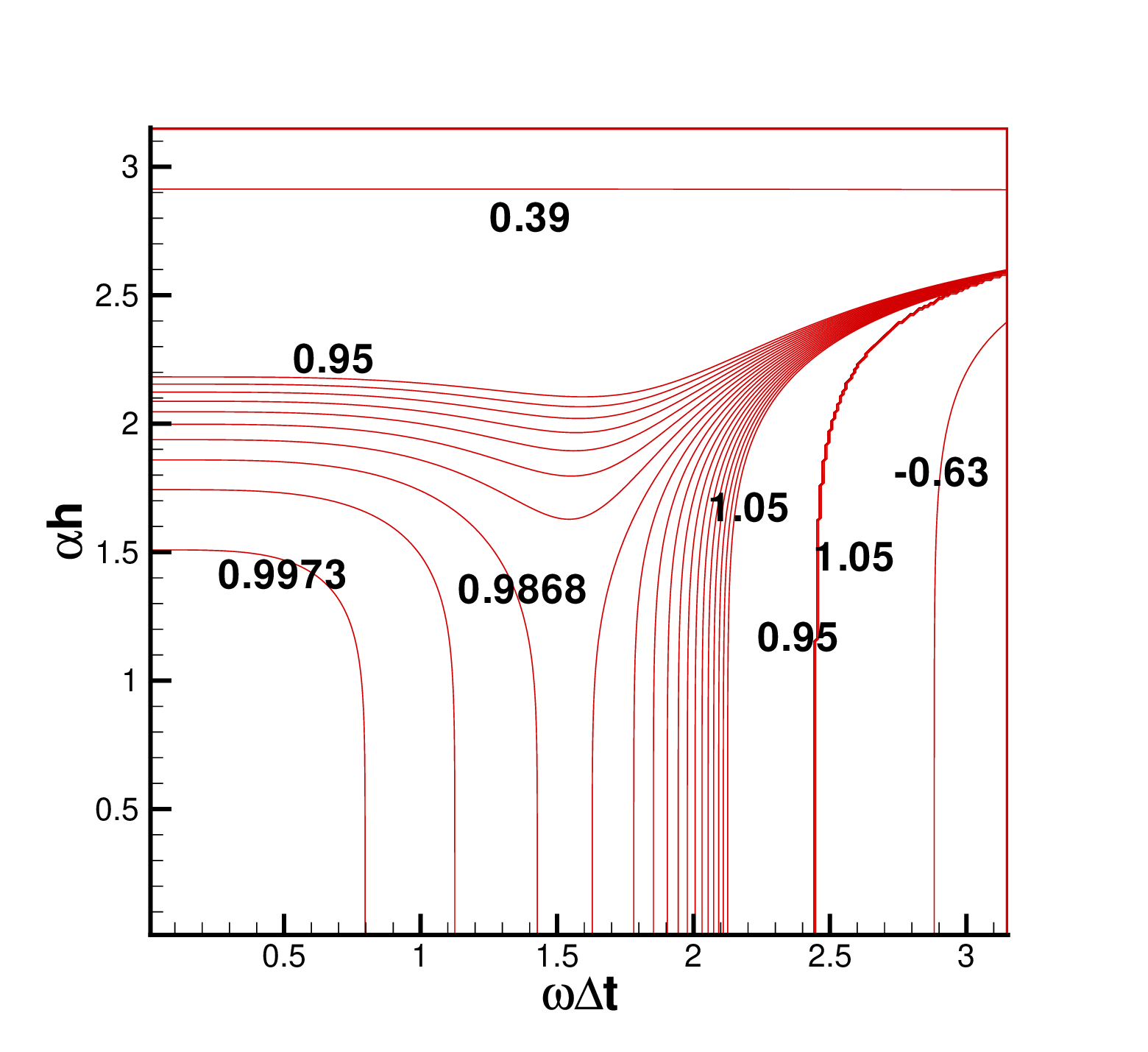}} }
\caption{Amplification factor (a) and (b), scaled numerical group speed (c) and (d), and
scaled numerical phase velocity (e) and (f) contours for RK4 time-integration scheme with:
(a)(c)(e) present DRPCCD5 and (b)(d)(f) CCD6 scheme \cite{bib:Chu140(1998)370-399}.}
\end{figure}

\begin{figure}[!hbp]
\vskip12pt
\centering
\mbox{\subfigure[]{\includegraphics[width=0.7\textwidth]{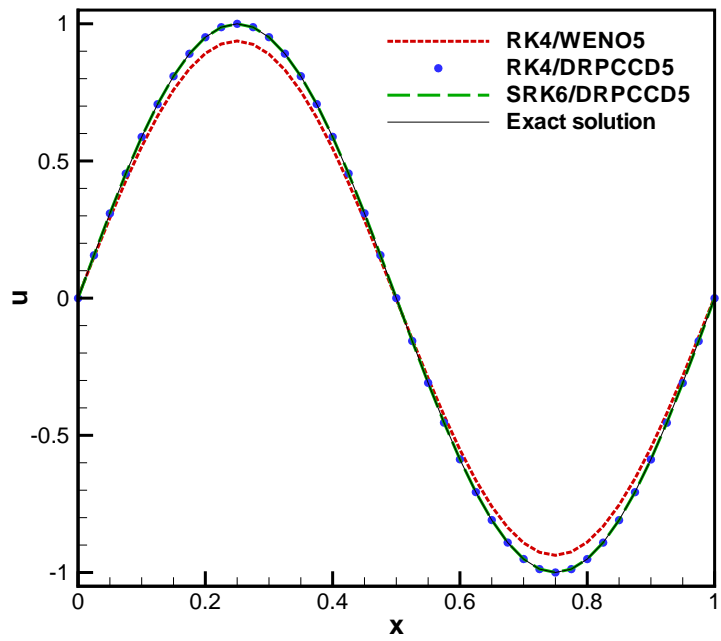}}}
\caption{The predicted results for linear advection problem$\#1$ are plotted using $40$ grids at $t=1000$.}
\end{figure}

\begin{figure}[!hbp]
\vskip12pt
\centering
\mbox{\subfigure[]{\includegraphics[width=0.7\textwidth]{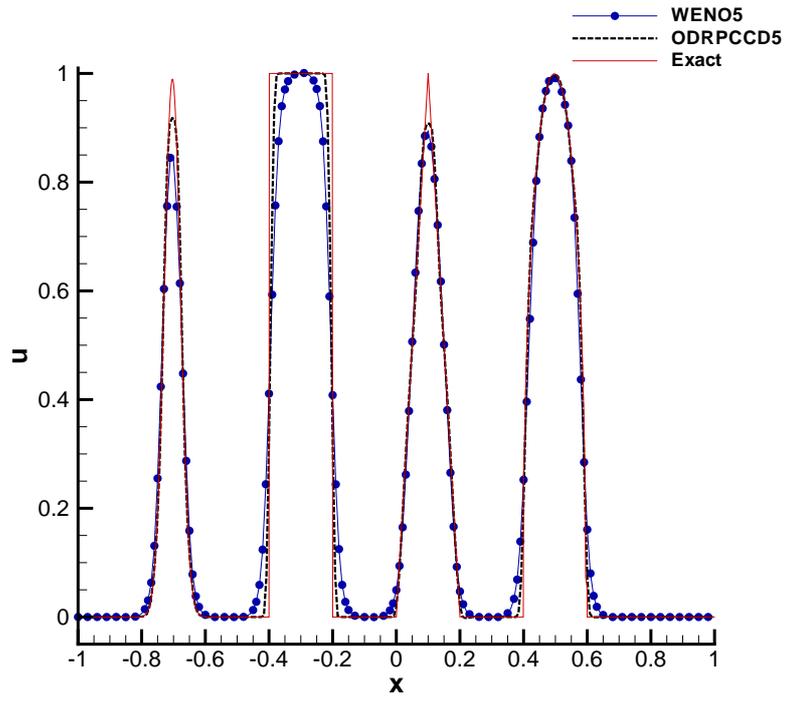}}}
\mbox{\subfigure[]{\includegraphics[width=0.7\textwidth]{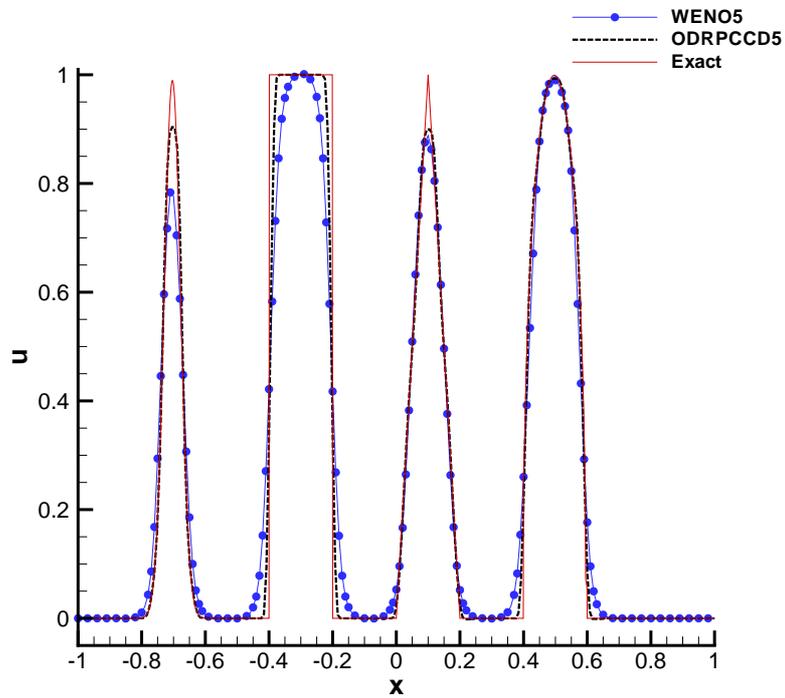}}}
\caption{The predicted results for linear advection problem$\#2$ are plotted at two different time
(a) $t=2$;
(b) $t=4$.}
\end{figure}

\begin{figure}[!hbp]
\vskip12pt
\centering
\mbox{\subfigure[]{\includegraphics[width=0.6\textwidth]{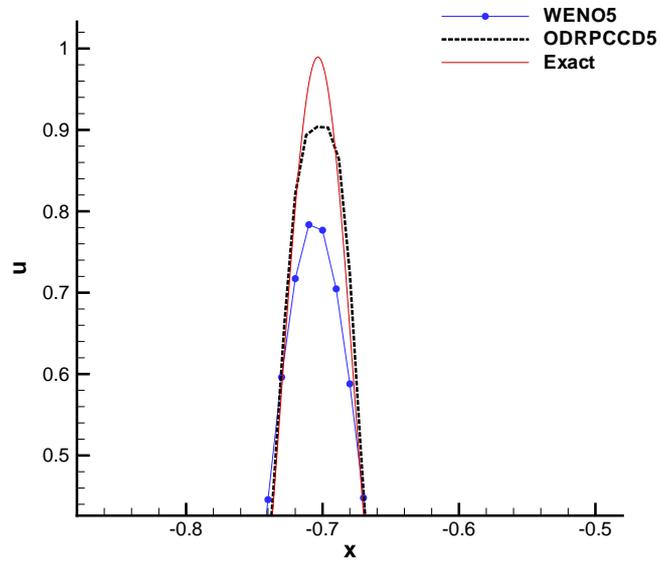}}}
\mbox{\subfigure[]{\includegraphics[width=0.6\textwidth]{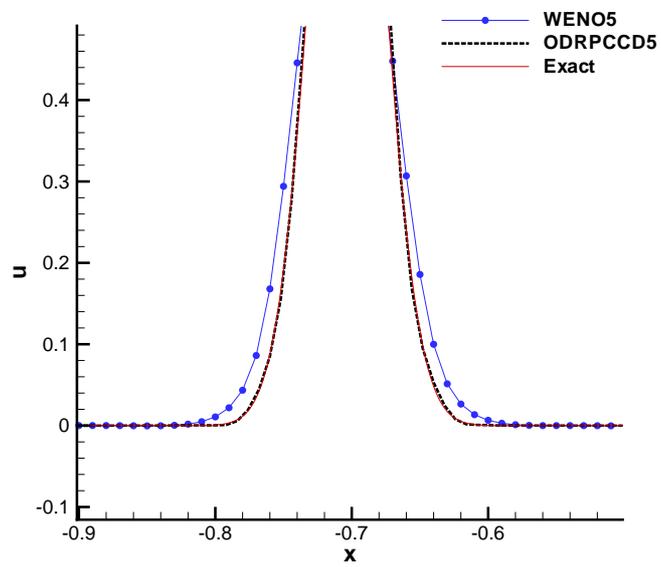}}}
\caption{Magnified solution for the exponential wave at $t=4$.
(a) Extreme;
(b) Bottom.}
\end{figure}

\begin{figure}[!hbp]
\vskip12pt
\centering
\mbox{\subfigure[]{\includegraphics[width=0.6\textwidth]{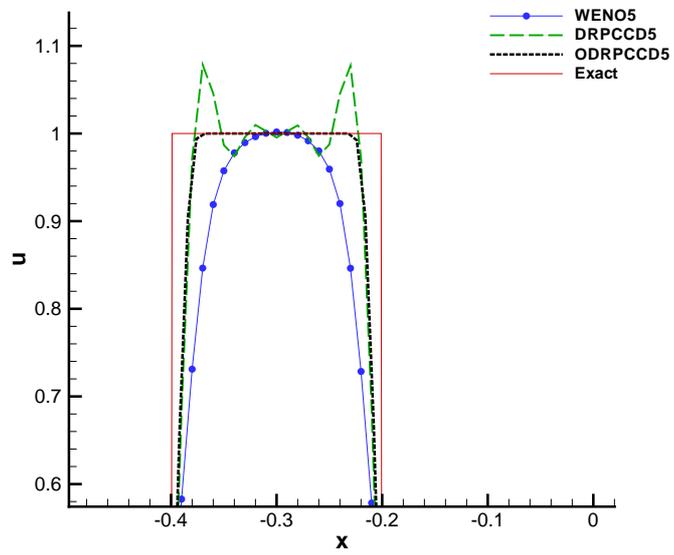}}}
\mbox{\subfigure[]{\includegraphics[width=0.6\textwidth]{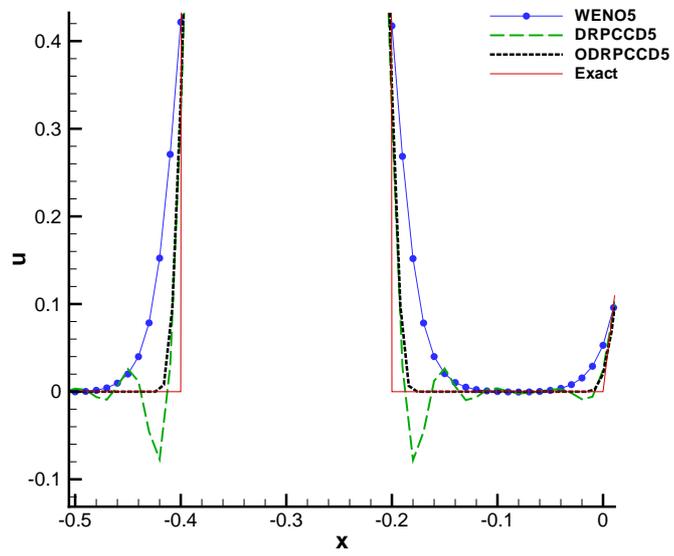}}}
\caption{Magnified solution for the square waves at $t=4$.
(a) Extreme;
(b) Bottom.}
\end{figure}

\begin{figure}[!hbp]
\vskip12pt
\centering
\mbox{\subfigure[]{\includegraphics[width=0.7\textwidth]{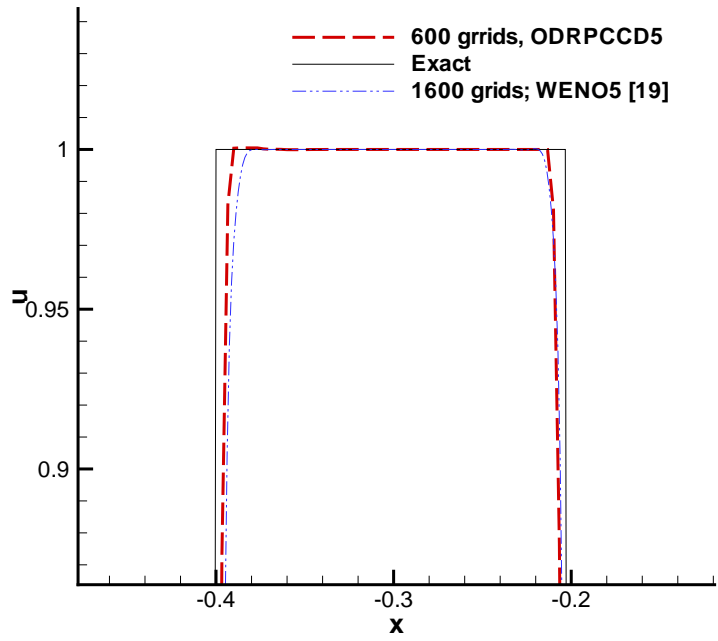}}}
\mbox{\subfigure[]{\includegraphics[width=0.7\textwidth]{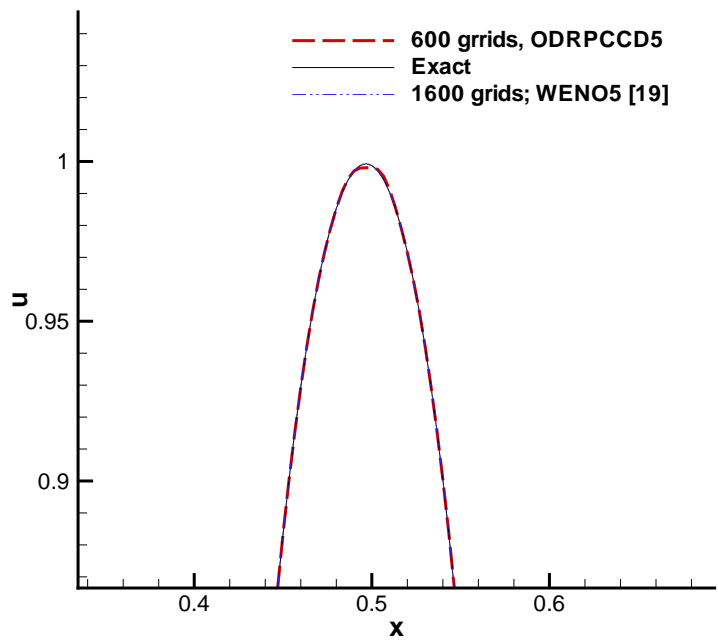}}}
\caption{The predicted results for linear advection problem$\#2$ are plotted at $t=4$.}
\end{figure}

%%%%%%%%%%%%%%%%%%%%%%%%%%%%%%%%%%%%%%%%%%%%%%%%%%%%%%%%%%%%%%%%%%%%%%%%%%%%%%%%%%%%%%%%%%%%%%%%%%%%%%%%%%
\begin{figure}[!hbp]
\vskip12pt
\centering
\mbox{\subfigure[]{\includegraphics[width=0.7\textwidth]{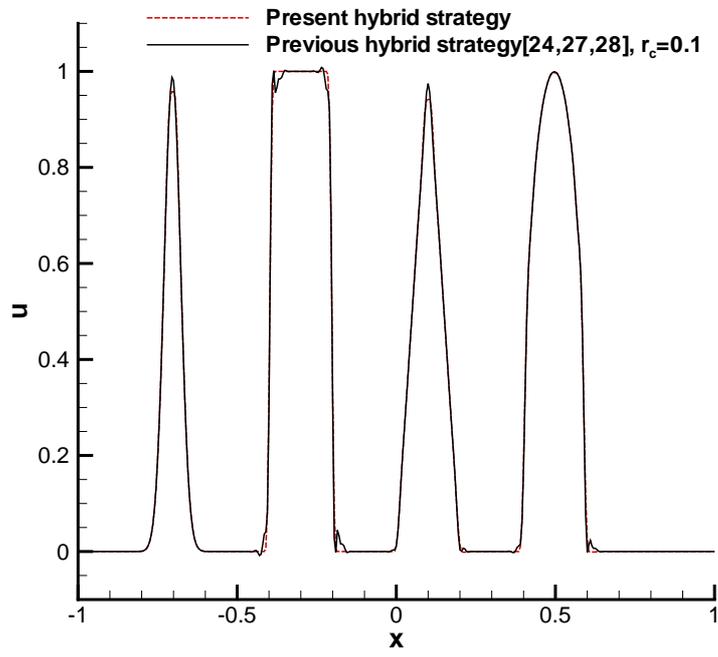}}}
\mbox{\subfigure[]{\includegraphics[width=0.7\textwidth]{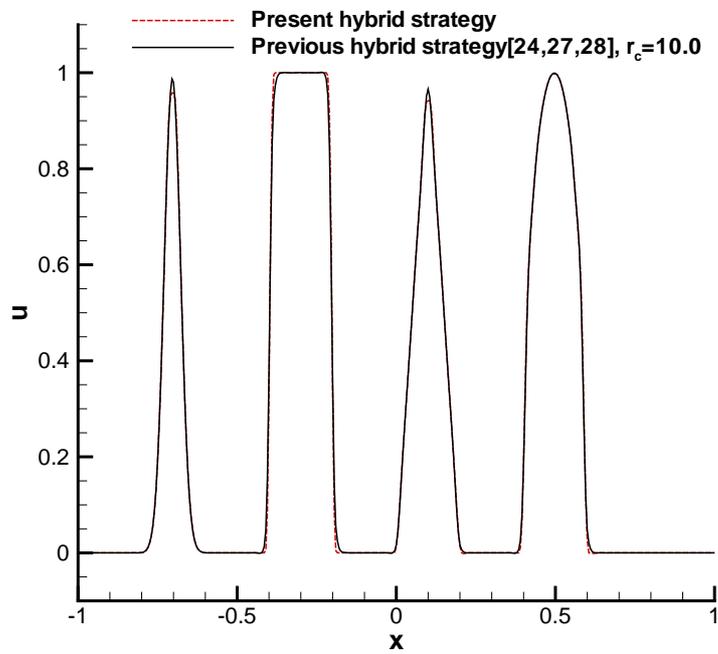}}}
\caption{Comparison of the results using the present hybrid strategy and the present hybrid strategy
(a) $r_{c}=0.1$; (b) $r_{c}=10.0$.}
\end{figure}

\begin{figure}[!hbp]
\vskip12pt
\centering
\mbox{\subfigure[]{\includegraphics[width=0.5\textwidth]{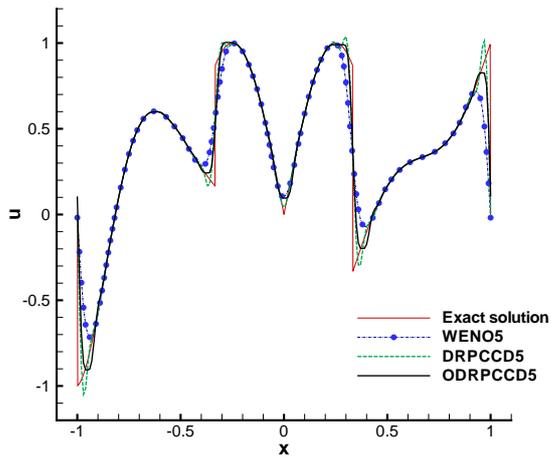}}
      \subfigure[]{\includegraphics[width=0.5\textwidth]{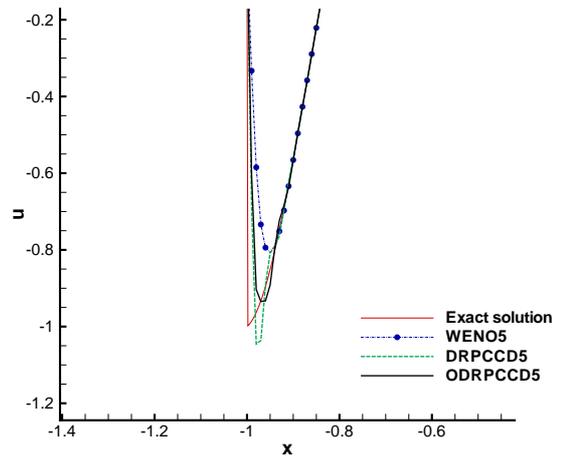}}}
\mbox{\subfigure[]{\includegraphics[width=0.5\textwidth]{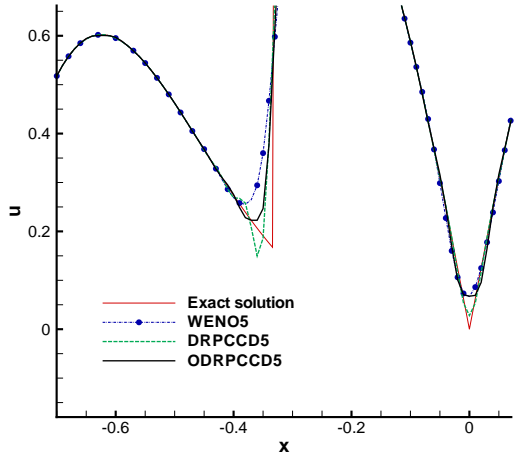}}
      \subfigure[]{\includegraphics[width=0.5\textwidth]{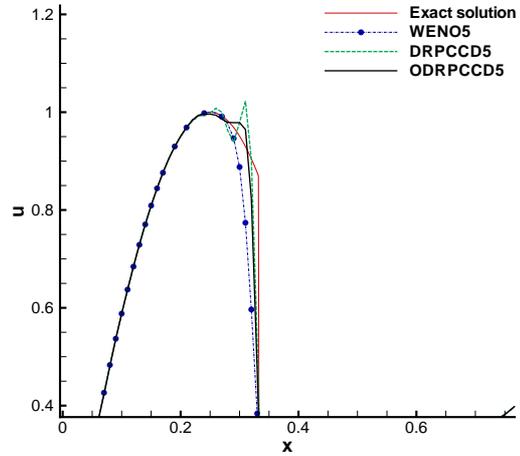}}}
\caption{(a) The predicted results for linear advection problem$\#3$ are plotted at $t=20$;
(b)Magnified solution between $-1.4\leq x \leq -0.4$; (c) Magnified solution between $-0.7\leq x \leq 0.1$; (d)Magnified solution between $0\leq x \leq 0.7$.}
\end{figure}

\begin{figure}[!hbp]
\vskip12pt
\centering
\mbox{\subfigure[]{\includegraphics[width=0.7\textwidth]{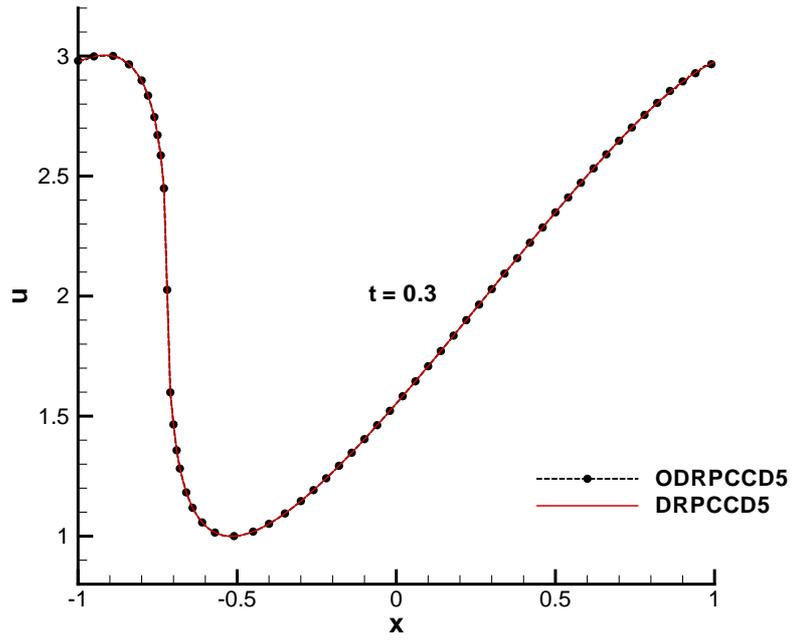}}}
\mbox{\subfigure[]{\includegraphics[width=0.7\textwidth]{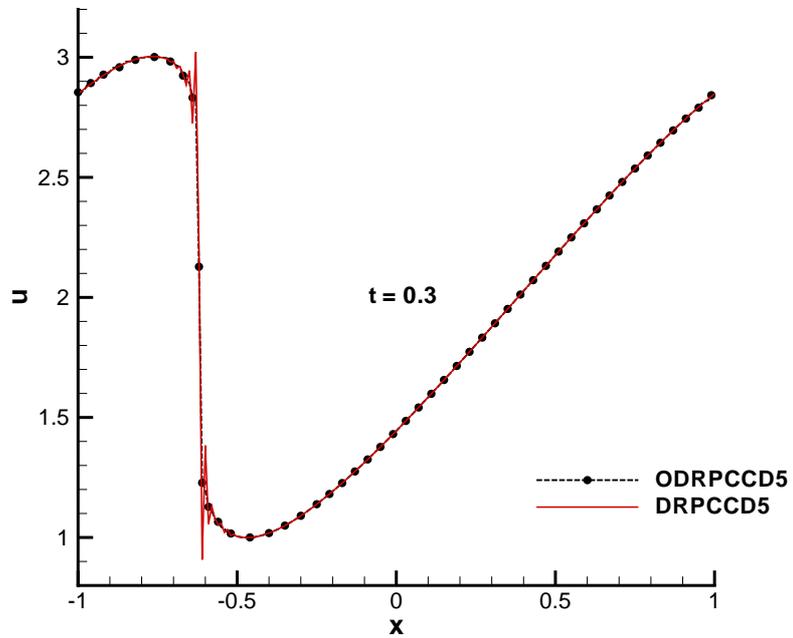}}}
\caption{The predicted results for non-linear advection problem are plotted at two different time.
(a) $t=0.3$;
(b) $t=0.35$.}
\end{figure}

%%%%%%%%%%%%%%%%%%%%%%%%%%%%%%%%%%%%%%%%%%%%%%%%%%%%%%%%%%%%%%%%%%%%%%%%%%%%%%%%%%%%%%%%%%%%%%%%%%%%%%%%%%
\begin{figure}[!hbp]
\vskip12pt
\centering
\mbox{\subfigure[]{\includegraphics[width=0.5\textwidth]{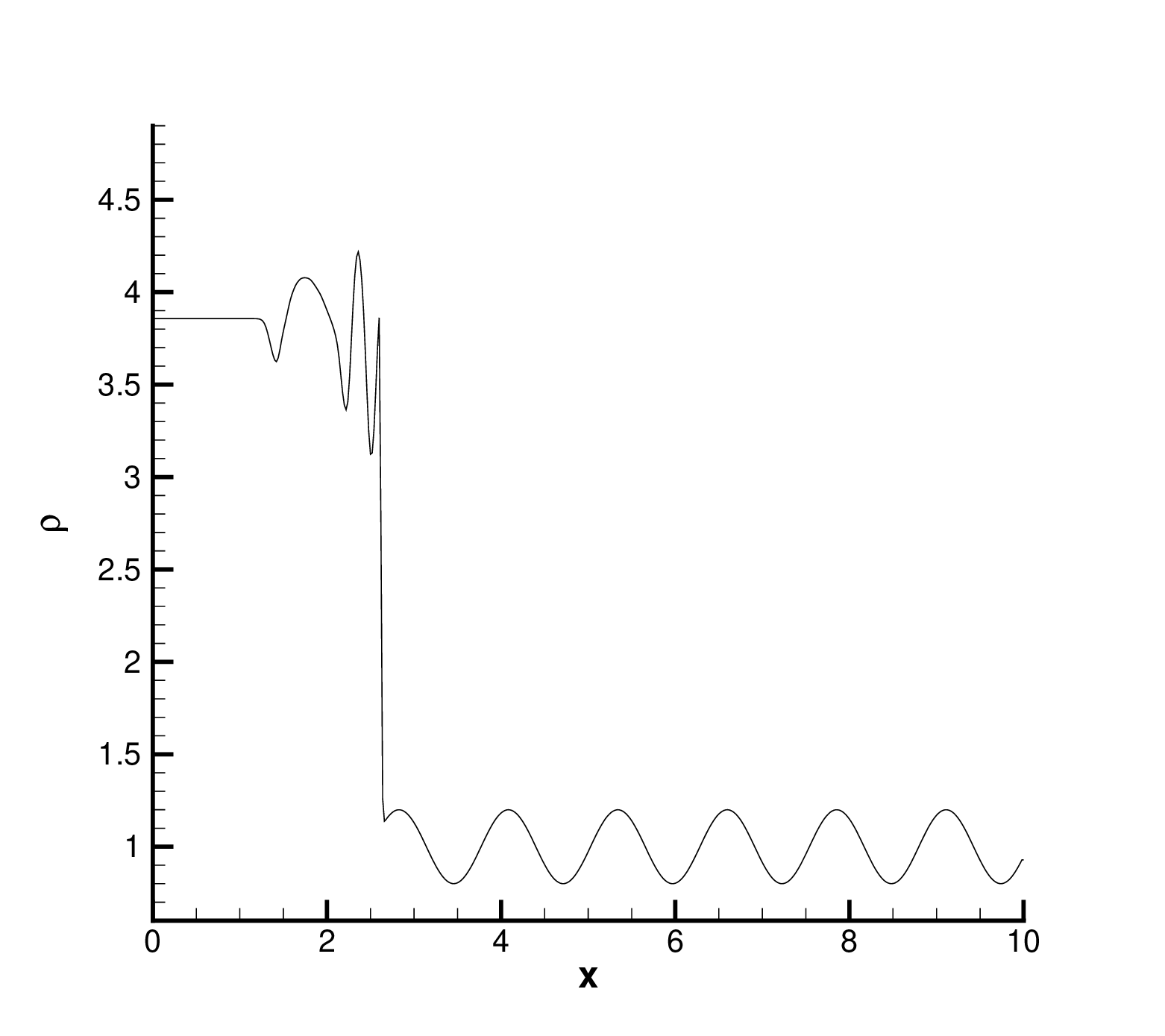}}
      \subfigure[]{\includegraphics[width=0.5\textwidth]{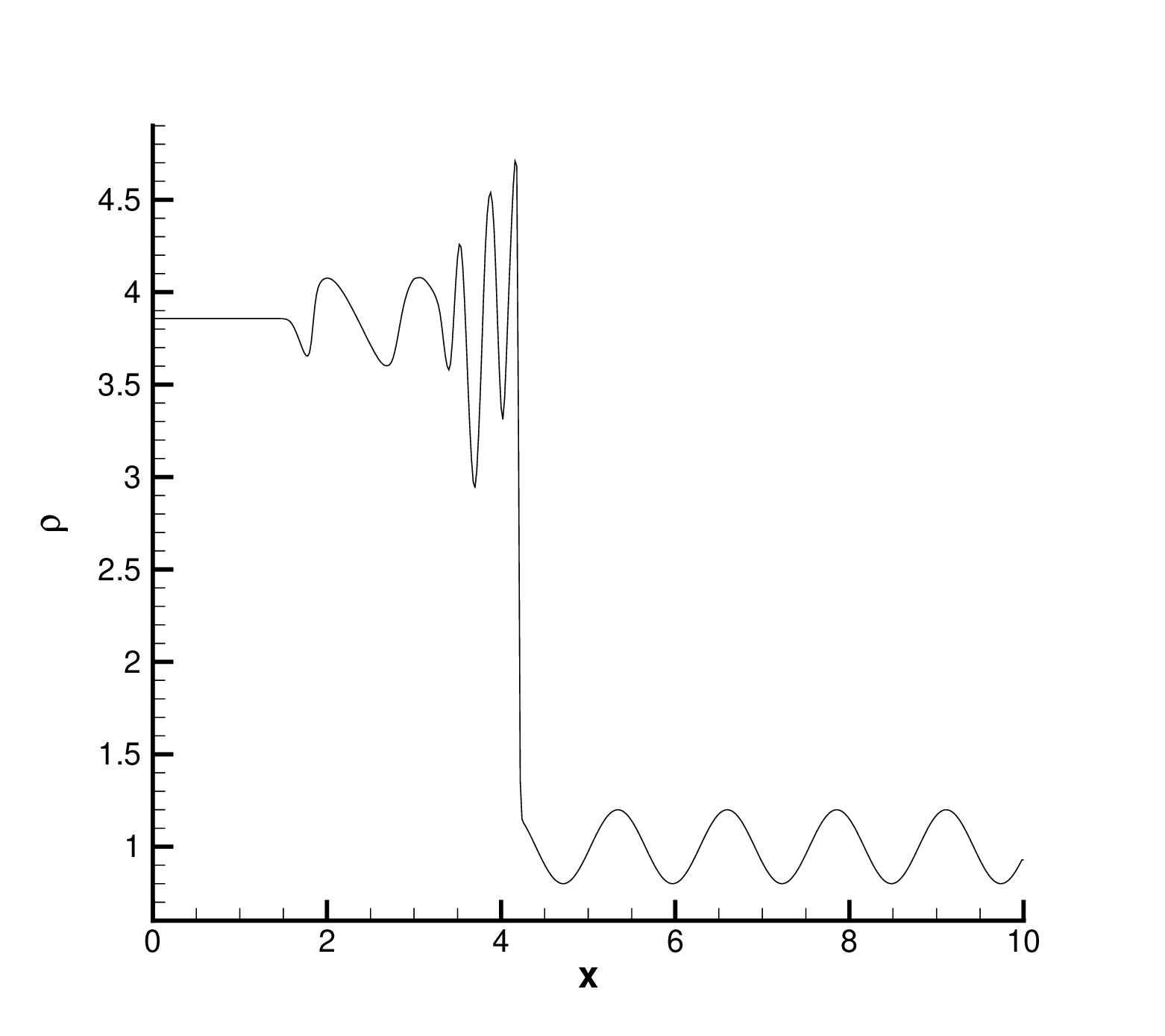}}}
\mbox{\subfigure[]{\includegraphics[width=0.5\textwidth]{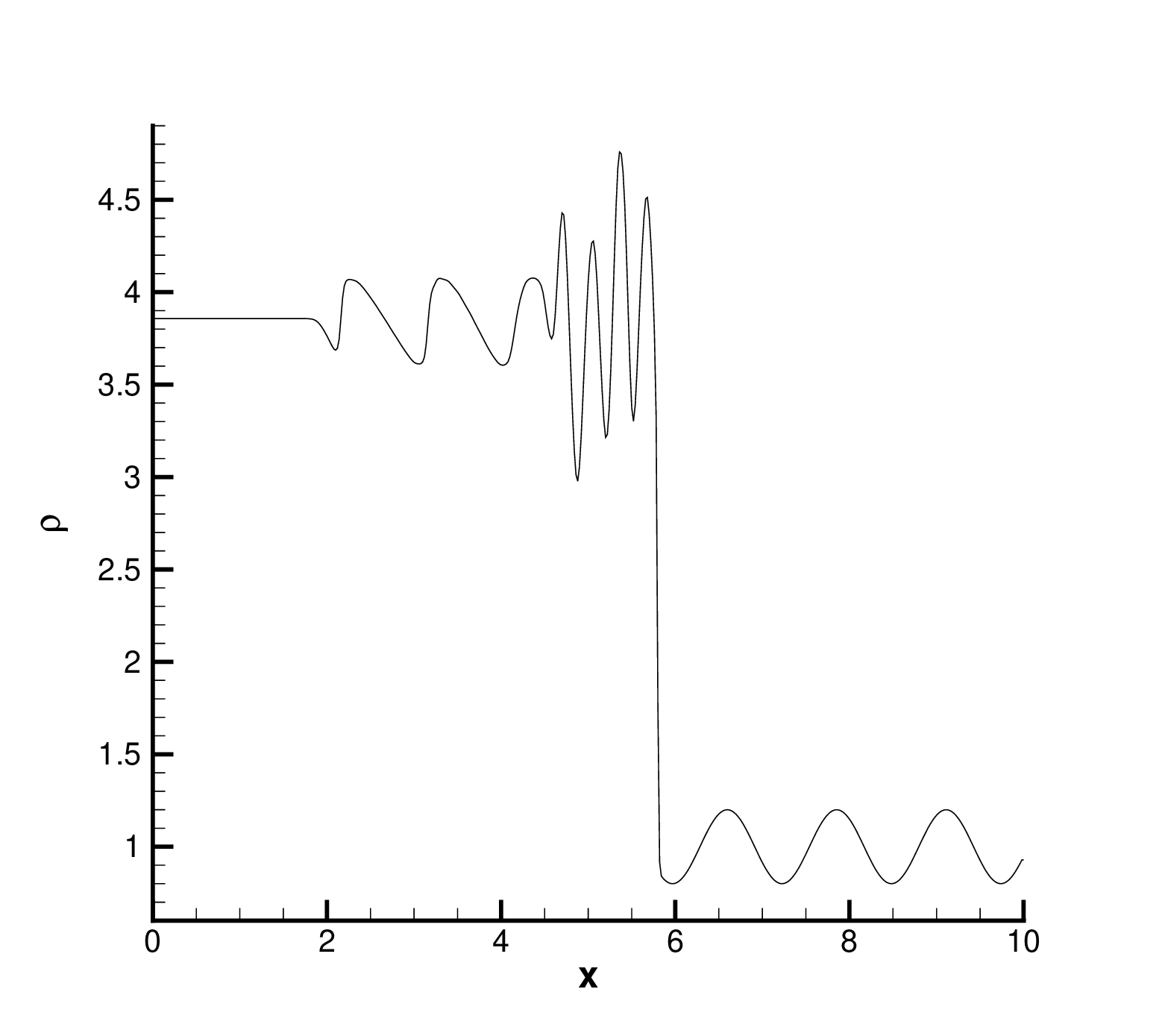}}
      \subfigure[]{\includegraphics[width=0.5\textwidth]{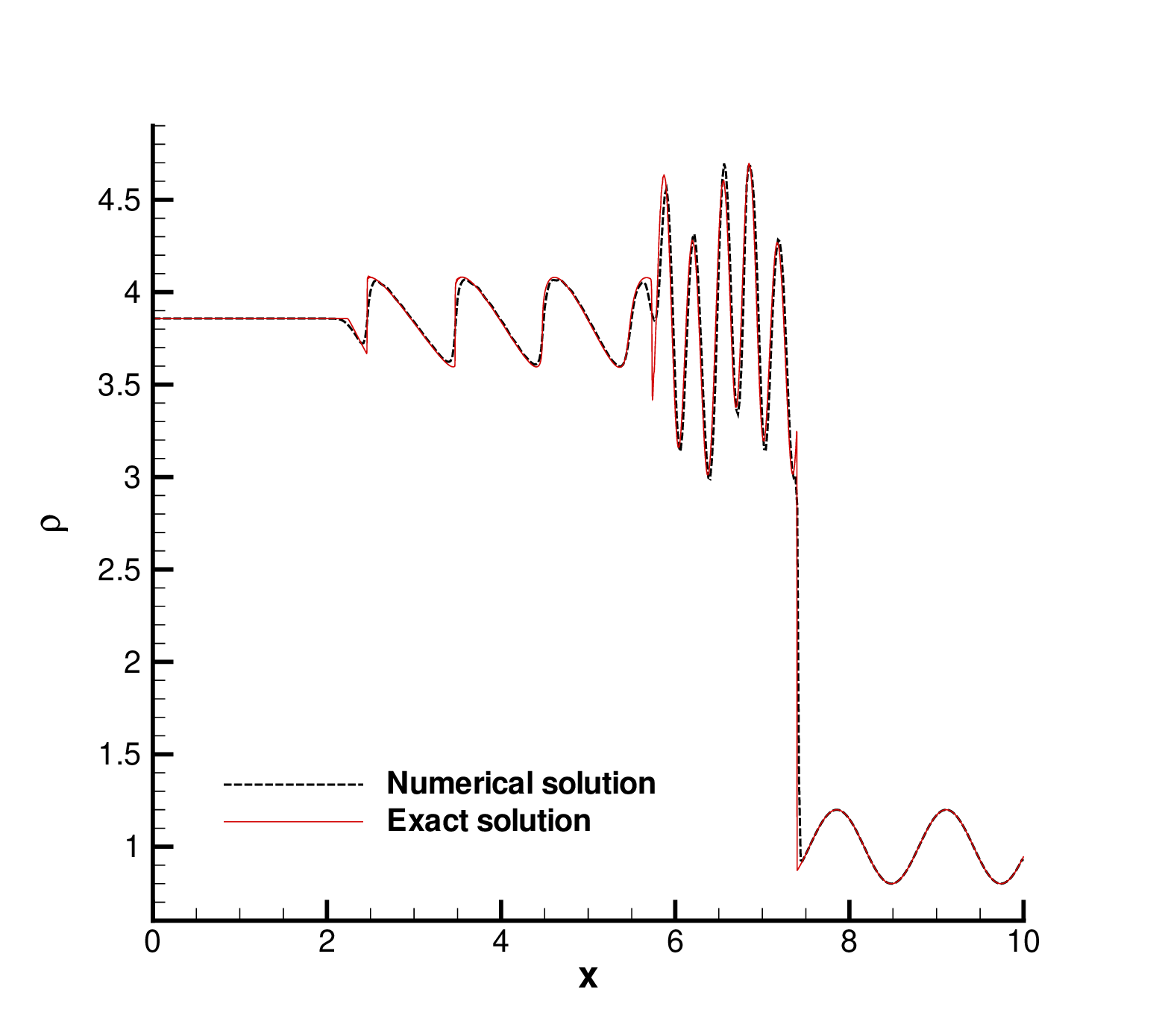}}}
\caption{The predicted results for the Shu-Osher problem are plotted at four different time.
(a) $t=0.45$;
(b) $t=0.9$;
(c) $t=1.35$;
(d) $t=1.8$.}
\end{figure}

%%%%%%%%%%%%%%%%%%%%%%%%%%%%%%%%%%%%%%%%%%%%%%%%%%%%%%%%%%%%%%%%%%%%%%%%%%%%%%%%%%%%%%%%%%%%%%%%%%%%%%%%%%
\begin{figure}[!hbp]
\vskip12pt
\centering
\mbox{\subfigure[]{\includegraphics[width=0.5\textwidth]{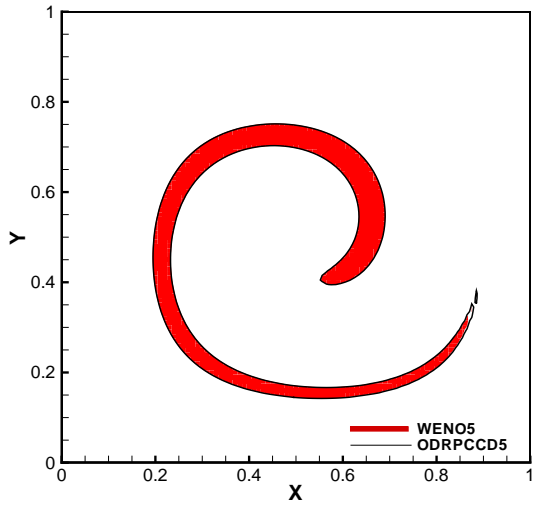}}
      \subfigure[]{\includegraphics[width=0.5\textwidth]{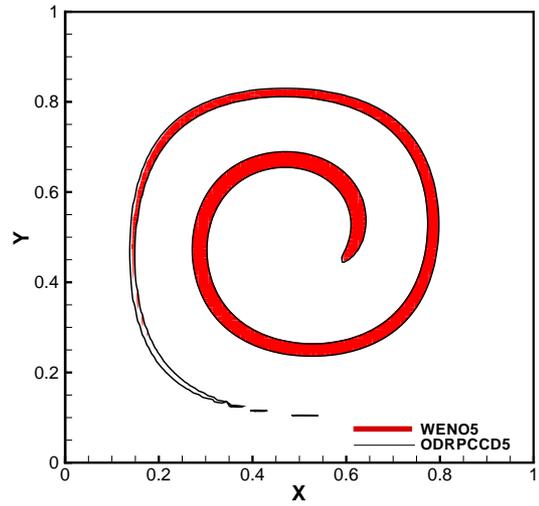}}}
\mbox{\subfigure[]{\includegraphics[width=0.5\textwidth]{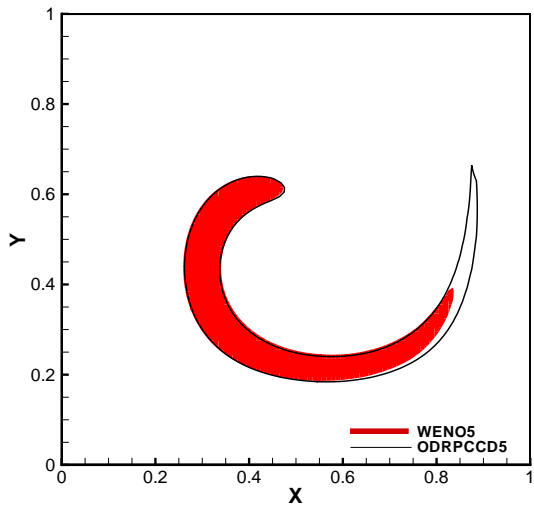}}
      \subfigure[]{\includegraphics[width=0.5\textwidth]{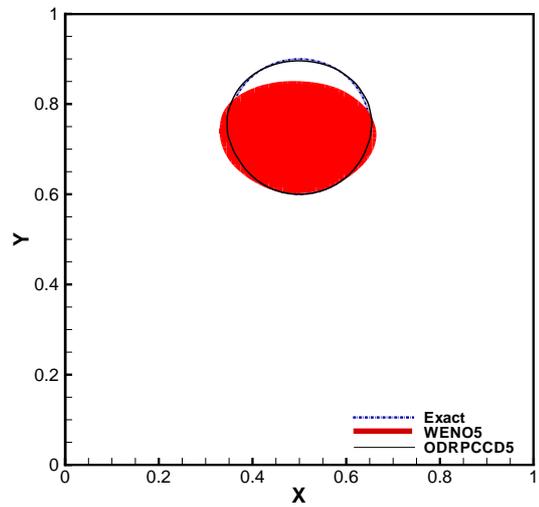}}}
\caption{Comparison of the results by ODRPCCD5 and WENO5 schemes for the vortex flow problem computed in $100\times100$ grids.
(a) $t=1.5$;
(b) $t=2.5$;
(c) $t=4$;
(d) $t=5$.}
\end{figure}

%%%%%%%%%%%%%%%%%%%%%%%%%%%%%%%%%%%%%%%%%%%%%%%%%%%%%%%%%%%%%%%%%%%%%%%%%%%%%%%%%%%%%%%%%%%%%%%%%%%%%%%%%%
\begin{figure}[!hbp]
\vskip12pt
\centering
\mbox{\subfigure[]{\includegraphics[width=0.7\textwidth]{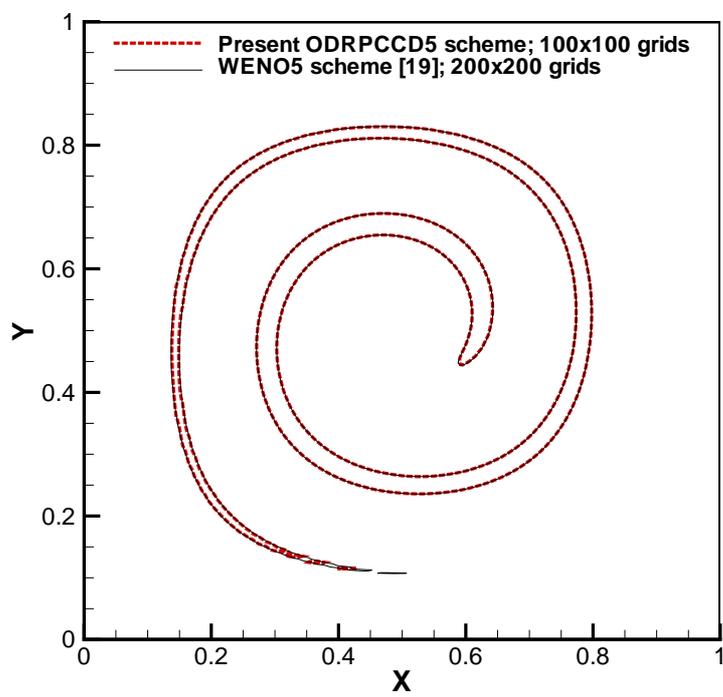}}}
\caption{Comparison of the results by ODRPCCD5 and WENO5 schemes at $t=2.5$.}
\end{figure}

%%%%%%%%%%%%%%%%%%%%%%%%%%%%%%%%%%%%%%%%%%%%%%%%%%%%%%%%%%%%%%%%%%%%%%%%%%%%%%%%%%%%%%%%%%%%%%%%%%%%%%%%%%
\begin{figure}[!hbp]
\vskip12pt
\centering
\mbox{\subfigure[]{\includegraphics[width=0.7\textwidth]{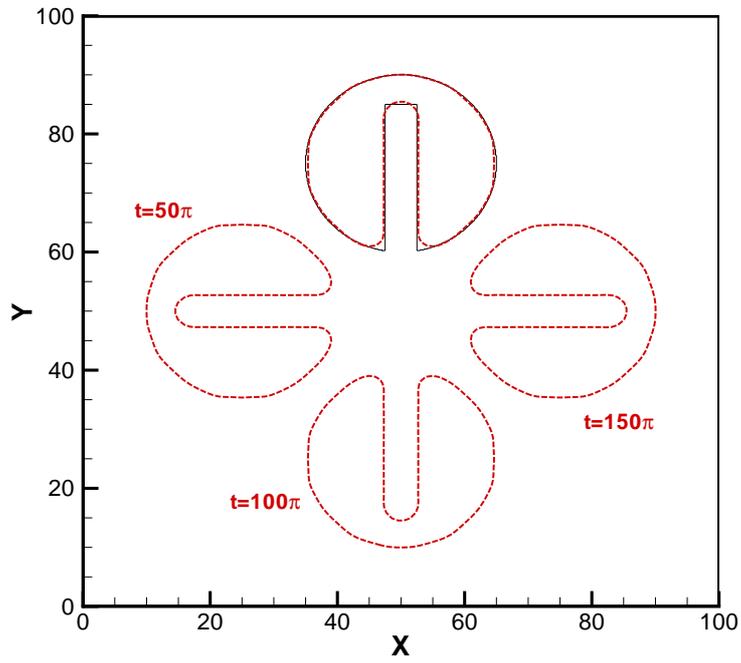}}}
\mbox{\subfigure[]{\includegraphics[width=0.7\textwidth]{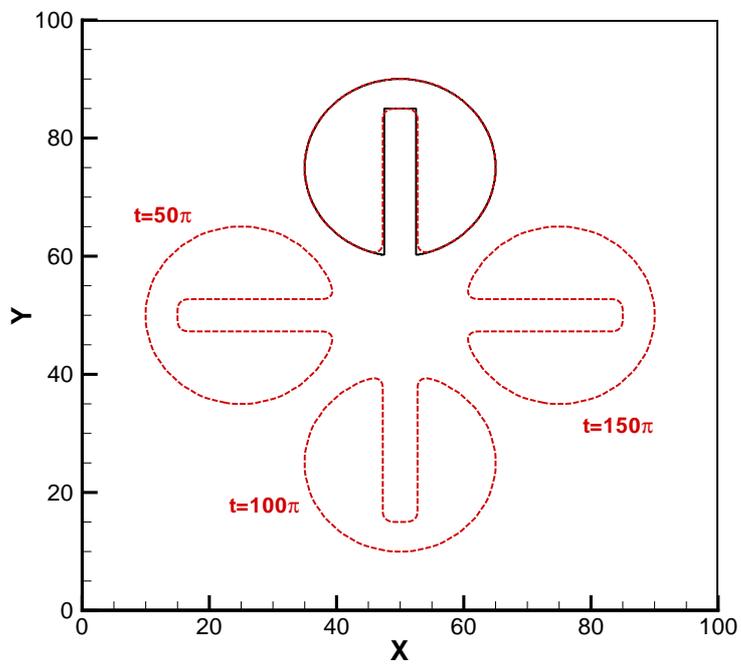}} }
\caption{The predicted results for the Zalesak's problem.
(a) $100\times100$ grids;
(b) $200\times200$ grids.}
\end{figure}

\end{document}